\documentclass{article}
\pdfoutput=1
\usepackage{lscape}
\usepackage{lipsum}
\usepackage{enumerate}
\usepackage{amsfonts}
\usepackage{amsmath}
\usepackage{comment}
\usepackage{textcomp}
\usepackage{amsthm, tikz}
\usepackage{amssymb}
\usepackage{color}
\usepackage{graphicx}
\usepackage{pdflscape}
\usepackage{afterpage}
\usepackage[utf8]{inputenc}
\usepackage[matrix,arrow,curve]{xy}
\usepackage{array}
\usepackage{mathtools}

\def\be{\begin{eqnarray}}
\def\ee{\end{eqnarray}}
\def\nn{\nonumber}

\usepackage{tikz}
\usepackage{braids}

\makeatletter
\renewcommand*{\@cite@ofmt}{\bfseries\hbox}
\makeatother

\def\l[{\phantom.[}

\hoffset= - 1.2in         

\begin{document}

\title{\vspace{0.1cm}{\Large {\bf Quantum Racah matrices and 3-strand braids in representation $[3,3]$}\vspace{.2cm}}
\author{
\ {\bf Sh.Shakirov$^{b}$}\footnote{shakirov.work@gmail.com},
\ \ and
 \ {\bf A.Sleptsov$^{a,b,c}$}\footnote{sleptsov@itep.ru}}
\date{ }
}

\maketitle

\vspace{-5.5cm}

\vspace{4.5cm}

\begin{center}
$^a$ {\small {\it ITEP, Moscow 117218, Russia}}\\
$^b$ {\small {\it Institute for Information Transmission Problems, Moscow 127994, Russia}}\\
$^c$ {\small {\it Moscow Institute of Physics and Technology, Dolgoprudny 141701, Russia}}

\end{center}

\vspace{.5cm}

\begin{abstract}
This paper is devoted to the advance in the project of systematic description of colored knot polynomials started in \cite{MMfing} -- explicit calculation of the {\it inclusive} Racah matrices for representation $R=[3,3]$. This is made possible by a powerful technique which we suggest in this paper -- the use of highest weight method in the basis of Gelfand-Tseitlin tables. Our result allows one to evaluate and investigate $[3,3]$-colored polynomials for arbitrary 3-strand knots, and this confirms many previous conjectures on various factorizations, universality, and differential expansions. Furthermore, with the help of a method developed in \cite{MMMStrick} we manage to calculate {\it exclusive} Racah matrices in $R=[3,3]$. Our results confirm a calculation of these matrices in \cite{Morozov33}, which was based on the conjecture of explicit form of differential expansion for twist knots. Explicit answers for Racah matrices and $[3,3]$-colored polynomials for 3-strand knots up to 10 crossings are available at \cite{knotebook}. With the help of our results for inclusive and exclusive Racah matrices, it is possible to compute $[3,3]$-colored HOMFLY-PT polynomial of any link for the so-called one-looped family links, which are obtained from arborescent links by adding one loop. 
\end{abstract}

\begin{center}
\end{center}

\vspace{.5cm}

\section{Introduction}

One of the central approaches in modern knot theory is the study of quantum invariants of knots and links. This approach originated in late 1980's from two complementary perspectives: first, Wilson loop observables in Chern-Simons quantum field theory \cite{CS}; second, intertwiners in representation theory of quantum groups \cite{RT}. Both approaches turned out to provide the same answer, associating to a knot $K$ a polynomial
\begin{align*}
K \ \ \ \mapsto \ \ \ H^{(K)}_{R}\big( q, a \big)
\end{align*}
\smallskip
in two variables $(q,a)$, called colored HOMFLY-PT knot polynomial. Among other things, this polynomial depends on a choice of color, i.e. a finite-dimensional representation $R$ of $SU(N)$ (it is possible to consider other Lie groups, but we focus our attention on the unitary case) where $a = q^N$. If $K$ is a link, then the HOMFLY-PT polynomial depends on several colors, associated to each connected component of the link.

The color dependence of knot or link polynomials turned out to be quite intricate. For simplest knots and links this dependence can be entirely tamed -- a celebrated example is the case of the double Hopf link, which consists of 3 linked unknots and gives rise to the topological vertex $C_{R_1,R_2,R_3}$ \cite{TopVertex}, one of the main technical tools of topological string theory on toric Calabi-Yau 3-folds. However, as soon as the knot or link becomes topologically non-trivial, complexity of the answer increases significantly. One may argue that such colored knot polynomials provide a topologically non-trivial generalization of WZW and Liouville conformal blocks \cite{KnotsVirasoro} and their color dependence should therefore be carefully studied. While only at its beginning, this study already revealed several interesting structures, such as traces of hidden integrability \cite{MMS} and the eigenvalue conjecture \cite{IMMMev}. As the available data continues to grow, one may expect many more structures to come to light.

While both the Chern-Simons path integral and quantum group representation theory allow \emph{in principle} to compute any desired colored knot polynomial, in practice representations as small as $[2,1]$ or $[3,1]$ have been out of reach until very recently. Inevitably, regardless of the approach one takes, one ends up with the necessity to deal with the so-called q-6j symbols, also known as quantum Racah-Wigner matrices

\be
U_{R_1,R_2,R_3}^{R_4}: \ \ \ \ \
\Big\{ (R_1\otimes R_2)\otimes R_3 \longrightarrow R_4\Big\} \ \ \longrightarrow \ \
\Big\{ R_1\otimes (R_2\otimes R_3) \longrightarrow R_4\Big\}
\ee
\smallskip\\
or, graphically,

\begin{center}
\includegraphics[width=0.5\textwidth]{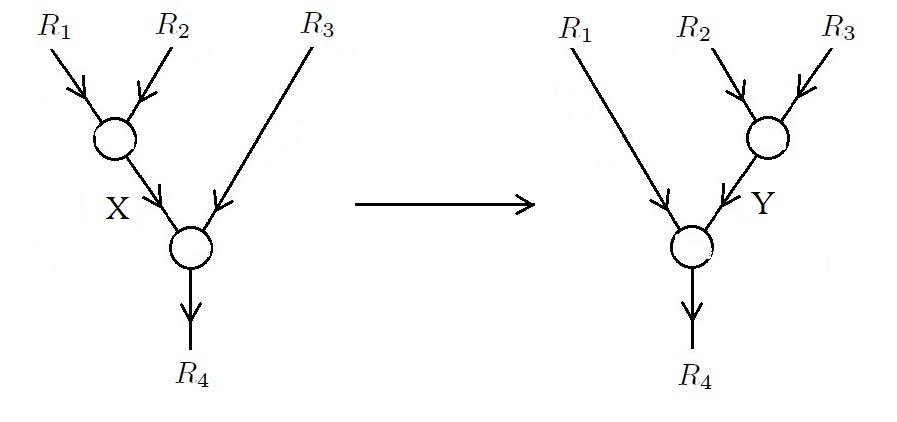}
\end{center}
The l.h.s. and r.h.s. represent two different bases in the space of intertwiners (invariant tensors of the quantum group) $R_1 \otimes R_2 \otimes R_3 \longrightarrow R_4$. Each basis is a collection of finitely many vectors, labeled by the choice of color in the internal channel ($X$ and $Y$ in the picture above) and the choice of intertwiner in each of the four trivalent vertices (white circles in the picture above). Racah matrix represents the change of basis from l.h.s. to r.h.s.

Depending on which problem in knot theory one addresses, one may need different types of Racah matrices. While in the original Reshetikhin-Turaev formalism one required all of them, some simplifications can be achieved by using its modern modifications \cite{modRT1, modRT2, modRTi}. In the course of this project we are particularly interested in the following two classes of Racah matrices,
\be
{\rm inclusive}: \ \ \ {\cal U}_Q\ \ {\rm with}\ \ R_1=R_2=R_3=R, \ \ \ \ R_4 = Q \in R^{\otimes 3}
\label{inc}
\ee
and
\be
{\rm exclusive}:\ \ \ S \  \ {\rm with} \ \ R_1=R_2=R_4=R, \ \ \ \ R_3 = \bar R \nn \\
\ \ \ \ \ \ {\rm  } \ \ \ \ \ \
\bar S \ \ {\rm with} \ \ R_1=R_3=R_4=R, \ \ \ \ R_2 = \bar R
\label{exc}
\ee
\smallskip\\
These two classes of Racah matrices allow to compute the HOMFLY-PT polynomials for two large classes of knots. The inclusive (the term refers to arbitrariness of the final representation $Q\in R^{\otimes 3}$) matrices ${\cal U}_Q$ allow to compute the $R$-colored HOMFLY polynomials for arbitrary 3-strand braids
${\cal L} = (m_1,n_1|m_2,n_2|\ldots)$ via \cite{MMMkn12},

\be
H_R^{(m_1,n_1|m_2,n_2|\ldots)}(A,q) =
\sum_{Q\in R^{\otimes 3}} \frac{d_Q}{d_R}\cdot
{\rm Tr}_Q \!\Big({\cal R}_Q^{m_1} {\cal U}_Q {\cal R}_Q^{n_1} {\cal U_Q}^\dagger
{\cal R}_Q^{m_2} {\cal U}_Q {\cal R}_Q^{n_2} {\cal U_Q}^\dagger\ldots \Big)
\label{3str}
\ee
\smallskip\\
where $d_R$ is the quantum dimension of representation $R$ for the Lie algebra $sl_N$,
expressed through the variable $A=q^N$, and ${\cal R}_Q$ is a diagonal matrix
with the entries

\be
\lambda_Y = \epsilon_Y q^{\varkappa_Y}, \ \ \ Y\in R^{\otimes 2}
\label{evY}
\ee
\smallskip\\
where $\varkappa_Y = \sum_{(i,j)\in Y} (i-j)$ is the value of Casimir operator in the representation $Y$, while $\epsilon_Y = \pm 1$ depending on whether $Y$ belongs to the symmetric or antisymmetric square of $R$. For other simple Lie algebras similar formulas exist, see \cite{MMuniv} for a short survey.

The exclusive matrices $S$ and $\bar S$, where only $R$ is picked up in the "final state" of the product $R\otimes R\otimes \bar R$, define \cite{MMMRS,MMMRSS} the building blocks ("fingers") which allow to compute the $R$-colored HOMFLY for arbitrary arborescent (double-fat) knots \cite{con,arbor} ${\cal K} = \{F^{I,k_I}\}$:

\be
H_R^{\{F_I\}} = \sum_{X_I \in R\otimes R \ {\rm or}\ R\otimes \bar R} \prod_{I,J} P_{X_I,X_J}
\prod_{k_I} F^{\{I,k_I\}}_{X_I}
\label{arbor}
\ee
\smallskip\\
where the propagators $P_{X'X''}$ connecting the vertices $I$ are just the matrices $S_{\bar X'X''}$ or $\bar S_{\bar X'\bar X''}$ (bars refer to the antiparallel rather than parallel double lines, the two parallel vertices never being connected), while the fingers attached to the vertices are arbitrary matrix elements of the type

\be
F_X =
\Big(\ldots S\bar {\cal R}^{l_3} S{\cal R}^{l_2} S^\dagger \bar{\cal R}^{l_1}\bar S\Big)_{\emptyset X}
\ee
\smallskip\\
Despite limited, these two classes of knots are quite rich and already quite informative: in particular, the second class contains non-trivial pairs of mutants. Computation of inclusive and exclusive Racah matrices can shed much light on the properties of these two classes of knots. 

The distinguishing of mutant knots is very complicated topological problem. It is known \cite{CromMort} that HOMFLY polynomials colored by multiplicity-free representations cannot distinguish mutants, while representations with multiplicities, at least in all cases known to us, can (see recent papers \cite{NRS_mut, Bishler1,Bishler2} on this topic). Representations without multiplicities are enumerated by all rectangular Young diagrams, so the representation $[3,3]$ is also of this type. However, quantum 6j-symbols are still of great interest for such representations. First of all, we are interested in their analytical properties. For example, it is known that any 6j-symbol in $SU(2)$  (and there are no multiplicities for $SU(2)$ at all) can be described with the help of Racah orthogonal polynomial, which is expressed in terms of the hypergeometric function $_4F_3$. From this expression, for example, it is easy to find all their $S_4\times S_3$ symmetries: the tetrahedron and Regge symmetries. Also, from the explicit expression, it is easy to obtain the famous asymptotic Ponzano-Regge formula \cite{PR} in terms of the volume of the corresponding tetrahedron. In papers \cite{Multicolored, AlekseevMS} it was found that inclusive and exclusive 6j-symbols with incoming symmetric representations of $SU(N)$ can be expressed in terms of $SU(2)$ ones, what allowed to find the counterpart of symmetries and asymptotic formula.

In addition, according to \cite{1loop1, 1loop2}, combining a 3-strand braid and arborescent braids \cite{arbor}, one can build up a large class of knots and links, which we call \textit{1-looped} links. This class definitely includes all arborescent links and the links that have 3-strand braid realizations. Note that these 3-strand and arborescent classes do not cover whole Rolfsen table, while the 1-looped class contains the entire Rolfsen table, all 3-strand links, all arborescent links, and, for example, all mutant knots with 11 intersections.

This paper represents yet another small step in this direction: we compute the inclusive and exclusive Racah matrices for $R = [3,3]$ using the highest weight method of \cite{MMMkn12}. Starting from 6 boxes, the method of \cite{MMMkn12} becomes increasingly complex; we cure this problem by passing from the earlier used tensor product bases to the Gelfand-Tsetlin basis \cite{GelfTset} for representations of quantum groups. This basis is distinguished from many perspectives -- practically, all computations become significantly faster -- and in our opinion deserves a separate publication to discuss. In this paper we do not expand on this topic, instead concentrating on results and applications in the particular case of $R = [3,3]$, which is an important special case and closes a number of open questions.

\section{Results for Racah matrices}

The best way to calculate inclusive quantum 6-j symbols is the highest weight method of \cite{MMMkn12}. With the help of this method iclusive Racah matrices were calculated for symmetric (and antisymmetric) representations $R=[r]$  (and $R=[1^r]$) up to level 5 in \cite{IMMMev}, for $R=[2,1]$ in \cite{MMMS21}, for $R=[3,1]$ in \cite{MMMS31} and for $R=[2,2]$ in \cite{MMMS22}. The method is straightforward but very tedious, especially for such representations $R$ which have multiplicities $R=[2,1]$ and $R=[3,1]$.

Exclusive quantum Racah matrices $S$ and $\bar S$ are known for symmetric (and antisymmetric) representations $R=[r]$ (and $R=[1^r]$) \cite{Racah,MMSpret}. Their general form was guessed after considering explicit formulas for several low levels. Actually, they look like a straightforward quantization and extension of the classical formulas cited in \cite{LL}. The only available at the moment non-rectangular diagram $R=[2,1]$ was calculated in \cite{GJ} by solving pentagon relations and additional specific symmetries of the representation $R=[2,1]$. The last available case $R=[2,2]$ (non-symmetric, but rectangular i.e. without multiplicities) was calculated in \cite{MMMStrick} with the help of a trick described in Section \ref{trick}. The problem of exclusive Racah matrices is that there is no good calculating method, all results were obtained by different methods, which actually are not working smoothly in other cases. The problem of the highest weight method is because conjugate representations depend on $N$ and thus the Racah matrices depend on $N$, hence, one needs to calculate matrices for various $N$ and reconstruct dependence on $N$ from a collection of the answers. A complexity of calculations grows very fast while $N$ increases.

\subsection{Specification to the case of $R=[3,3]$}

This case is relatively simple, because it does not contain multiplicities, the number of matrices not so big and their sizes are rather small,  still this is a new piece of knowledge.

Decomposition of the square
{\small
\be
[3,3]\otimes [3,3] = [6, 6]\oplus\,[6, 5, 1]\oplus\,[6, 4, 2]\oplus\,[6, 3, 3]\oplus\,[5, 5, 1, 1]\oplus\,[5, 4, 2, 1]\oplus\,[5, 3, 3, 1]\oplus\,[4, 4, 2, 2]   \oplus\,[4, 3, 3, 2] \oplus\,[3, 3, 3, 3]
\ee
}
contains ten irreducible representations.

The maximal size of the Racah mixing matrices will be $12\times 12$. All representations come with no multiplicities, but not all ${\cal R}$-matrix are different in contrast to $[2,2]$ case. For example, ${\cal R}$-matrices of $[8,5,2,2,1]$ and $[8,4,4,1,1]$ are difened by same intermediate diagrams $[6,4,2]$ and $[5,4,2,1]$. Another fact is that all eigenvalues of ${\cal R}$-matrices are different (there are no accidental coincidences, which are often encountered for non-rectangular representations $R$). These facts give a chance that the mixing matrices can be defined from eigenvalue hypothesis \cite{IMMMev}, which provides explicit expressions for the entries of Racah matrices through eigenvalues of the ${\cal R}$-matrices (they are given in \cite{IMMMev} for sizes up to $5$ and size $6$ was later described in \cite{MMuniv}).

Representation content of the cube is
{\footnotesize
\be
[3,3]\otimes\Big([3,3]\otimes[3,3]\Big) &=&
[9, 9]+[6, 6, 6]+2[7, 6, 5]+[7, 7, 4]+[8, 5, 5]+3[8, 6, 4]+2[8, 7, 3]+[8, 8, 2]+2[9, 5, 4]+4[9, 6, 3]+\nn \\
&+&3[9, 7, 2]+2[9, 8, 1]+3[5, 5, 4, 4]+[5, 5, 5, 3]+[6, 4, 4, 4]+8[6, 5, 4, 3]+3[6, 5, 5, 2]+10[6, 6, 3, 3]+\nn \\
&+&6[6, 6, 4, 2]+3[6, 6, 5, 1]+3[7, 4, 4, 3]+6[7, 5, 3, 3]+9[7, 5, 4, 2]+3[7, 5, 5, 1]+12[7, 6, 3, 2]+\nn \\
&+&6[7, 6, 4, 1]+6[7, 7, 2, 2]+3[7, 7, 3, 1]+3[8, 4, 3, 3]+3[8, 4, 4, 2]+6[8, 5, 3, 2]+6[8, 5, 4, 1]+\nn \\
&+&3[8, 6, 2, 2]+9[8, 6, 3, 1]+6[8, 7, 2, 1]+3[8, 8, 1, 1]+[9, 3, 3, 3]+2[9, 4, 3, 2]+[9, 4, 4, 1]+\nn \\
&+&[9, 5, 2, 2]+3[9, 5, 3, 1]+2[9, 6, 2, 1]+[9, 7, 1, 1]+2[5, 4, 3, 3, 3]+4[5, 4, 4, 3, 2]+2[5, 4, 4, 4, 1]+\nn \\
&+&3[5, 5, 3, 3, 2]+3[5, 5, 4, 2, 2]+6[5, 5, 4, 3, 1]+2[5, 5, 5, 2, 1]+4[6, 3, 3, 3, 3]+9[6, 4, 3, 3, 2]+\nn \\
&+&5[6, 4, 4, 2, 2]+6[6, 4, 4, 3, 1]+6[6, 5, 3, 2, 2]+12[6, 5, 3, 3, 1]+10[6, 5, 4, 2, 1]+3[6, 5, 5, 1, 1]+\nn \\
&+&[6, 6, 2, 2, 2]+8[6, 6, 3, 2, 1]+3[6, 6, 4, 1, 1]+3[7, 3, 3, 3, 2]+6[7, 4, 3, 2, 2]+6[7, 4, 3, 3, 1]+\nn \\
&+&6[7, 4, 4, 2, 1]+3[7, 5, 2, 2, 2]+12[7, 5, 3, 2, 1]+6[7, 5, 4, 1, 1]+6[7, 6, 2, 2, 1]+6[7, 6, 3, 1, 1]+\nn \\
&+&3[7, 7, 2, 1, 1]+2[8, 3, 3, 3, 1]+4[8, 4, 3, 2, 1]+2[8, 4, 4, 1, 1]+2[8, 5, 2, 2, 1]+6[8, 5, 3, 1, 1]+\nn \\
&+&4[8, 6, 2, 1, 1]+2[8, 7, 1, 1, 1]+[3, 3, 3, 3, 3, 3]+2[4, 3, 3, 3, 3, 2]+3[4, 4, 3, 3, 2, 2]+[4, 4, 3, 3, 3, 1]+\nn \\
&+&[4, 4, 4, 2, 2, 2]+2[4, 4, 4, 3, 2, 1]+[4, 4, 4, 4, 1, 1]+[5, 3, 3, 3, 2, 2]+3[5, 3, 3, 3, 3, 1]+2[5, 4, 3, 2, 2, 2]+\nn \\
&+&6[5, 4, 3, 3, 2, 1]+3[5, 4, 4, 2, 2, 1]+3[5, 4, 4, 3, 1, 1]+[5, 5, 2, 2, 2, 2]+3[5, 5, 3, 2, 2, 1]+6[5, 5, 3, 3, 1, 1]+\nn \\
&+&3[5, 5, 4, 2, 1, 1]+[5, 5, 5, 1, 1, 1]+2[6, 3, 3, 3, 2, 1]+4[6, 4, 3, 2, 2, 1]+3[6, 4, 3, 3, 1, 1]+3[6, 4, 4, 2, 1, 1]+\nn \\
&+&2[6, 5, 2, 2, 2, 1]+6[6, 5, 3, 2, 1, 1]+2[6, 5, 4, 1, 1, 1]+3[6, 6, 2, 2, 1, 1]+[6, 6, 3, 1, 1, 1]+[7, 3, 3, 3, 1, 1]+\nn \\
&+&2[7, 4, 3, 2, 1, 1]+[7, 4, 4, 1, 1, 1]+[7, 5, 2, 2, 1, 1]+3[7, 5, 3, 1, 1, 1]+2[7, 6, 2, 1, 1, 1]+[7, 7, 1, 1, 1, 1]
\nn
\ee
}
and the inclusive Racah matrices form the collection

{\scriptsize
\be
\begin{array}{|c|p{14cm}|c|}
\hline
&&\text{number of}\\
\text{matrix size} & \hspace{6.2cm} Q &  \\
&&\text{matrices}\\ \hline && \\
1 & [9, 9], [9, 7, 1, 1], [9, 5, 2, 2], [9, 4, 4, 1], [9, 3, 3, 3], [8, 8, 2], [8, 5, 5], [7, 7, 4], [7, 7, 1, 1, 1, 1], [7, 5, 2, 2, 1, 1], [7, 4, 4, 1, 1, 1], [7, 3, 3, 3, 1, 1], [6, 6, 6], [6, 4, 4, 4], [6, 6, 2, 2, 2], [6, 6, 3, 1, 1, 1], [5, 5, 5, 3], [5, 5, 5, 1, 1, 1], [5, 5, 2, 2, 2, 2], [5, 3, 3, 3, 2, 2], [4, 4, 4, 4, 1, 1], [4, 4, 4, 2, 2, 2], [4, 4, 3, 3, 3, 1], [3, 3, 3, 3, 3, 3] & 24 \\
&&\\ \hline && \\
2 & [9, 8, 1], [9, 5, 4], [9, 6, 2, 1], [9, 4, 3, 2], [8, 7, 3], [8, 7, 1, 1, 1], [8, 5, 2, 2, 1], [8, 4, 4, 1, 1], [8, 3, 3, 3, 1], [7, 6, 5], [7, 6, 2, 1, 1, 1], [7, 4, 3, 2, 1, 1], [6, 5, 4, 1, 1, 1], [6, 5, 2, 2, 2, 1], [6, 3, 3, 3, 2, 1], [5, 5, 5, 2, 1], [5, 4, 4, 4, 1], [5, 4, 3, 3, 3], [5, 4, 3, 2, 2, 2], [4, 4, 4, 3, 2, 1], [4, 3, 3, 3, 3, 2] & 21 \\
&&\\ \hline && \\
3 & [9, 7, 2], [9, 5, 3, 1], [8, 6, 4], [8, 8, 1, 1], [8, 6, 2, 2], [8, 4, 4, 2], [8, 4, 3, 3], [7, 7, 3, 1], [7, 5, 5, 1], [7, 4, 4, 3], [7, 7, 2, 1, 1], [7, 5, 2, 2, 2], [7, 3, 3, 3, 2], [7, 5, 3, 1, 1, 1], [6, 6, 5, 1], [6, 5, 5, 2], [6, 6, 4, 1, 1], [6, 5, 5, 1, 1], [6, 6, 2, 2, 1, 1], [6, 4, 4, 2, 1, 1], [6, 4, 3, 3, 1, 1], [5, 5, 4, 4], [5, 5, 4, 2, 2], [5, 5, 3, 3, 2], [5, 5, 4, 2, 1, 1], [5, 5, 3, 2, 2, 1], [5, 4, 4, 3, 1, 1], [5, 4, 4, 2, 2, 1], [5, 3, 3, 3, 3, 1], [4, 4, 3, 3, 2, 2] & 30 \\
&&\\ \hline && \\
4 & [9, 6, 3], [8, 6, 2, 1, 1], [8, 4, 3, 2, 1], [6, 3, 3, 3, 3], [6, 4, 3, 2, 2, 1], [5, 4, 4, 3, 2] & 6 \\
&&\\ \hline && \\
5 & [6, 4, 4, 2, 2] & 1 \\
&&\\ \hline && \\
6 & [8, 7, 2, 1], [8, 5, 4, 1], [8, 5, 3, 2], [8, 5, 3, 1, 1], [7, 7, 2, 2], [7, 6, 4, 1], [7, 5, 3, 3], [7, 6, 3, 1, 1], [7, 6, 2, 2, 1], [7, 5, 4, 1, 1], [7, 4, 4, 2, 1], [7, 4, 3, 3, 1], [7, 4, 3, 2, 2], [6, 6, 4, 2], [6, 5, 3, 2, 2], [6, 4, 4, 3, 1], [6, 5, 3, 2, 1, 1], [5, 5, 4, 3, 1], [5, 5, 3, 3, 1, 1], [5, 4, 3, 3, 2, 1] & 20 \\
&&\\ \hline && \\
7 & -- & 0 \\
&&\\ \hline && \\
8 & [6, 5, 4, 3], [6, 6, 3, 2, 1] & 2 \\
&&\\ \hline && \\
9 & [8, 6, 3, 1], [7, 5, 4, 2], [6, 4, 3, 3, 2] & 3 \\
&&\\ \hline && \\
10 & [6, 6, 3, 3], [6, 5, 4, 2, 1] & 2 \\
&&\\ \hline && \\
11 & -- & 0 \\
&&\\ \hline && \\
12 & [7, 6, 3, 2], [7, 5, 3, 2, 1], [6, 5, 3, 3, 1] & 3 \\
&&\\ \hline
 \end{array}
 \nn
\ee
}
All matrices were calculated with the help of the highest weight method using Gelfand-Tseitlin basis. Results are available online at \cite{knotebook}. All matrices of size up to four are nicely handled by the eigenvalue hypothesis, but we checked them by the direct calculations. Here we list one Racah matrix for $Q=[7,6,3,2]$ as an illustration (factors $u_i^j$ are listed explicitly in Appendix A below).

\pagebreak

\begin{landscape}
{\fontsize{3pt}{8pt}{
$\vspace{2cm}$
$\hspace{-7mm}$
$
\arraycolsep=0.1em
\hspace{-0.6cm}\left(\hspace{-0.5mm}\begin{array}{cccccccccccc}
\rule[-.3\baselineskip]{0pt}{6ex} \frac{[3]^2[2]}{[7][6]^2[5]} u_{1}^{1}  &  \sqrt{\frac{[10][4]!}{[7][6][5]}} u_{1}^{2}  &  \frac{[9][3][2]}{[6]^2\sqrt{[5]}} u_{1}^{3}  &  \sqrt{\frac{[10][4][3]^3}{[7][5]}} u_{1}^{4}  &  \frac{[9][3]\sqrt{[3]}}{[6][5]} u_{1}^{5}  &  \frac{[3]}{\sqrt{[7]}[6]} u_{1}^{6}  &  \sqrt{[10][4]}\frac{[4]![3]!}{[7]!} u_{1}^{7}  &  \sqrt{[10][9][3]!}\frac{[5][3]}{[6]^2} u_{1}^{8}  &  \sqrt{\frac{[10]}{[6]!}}\frac{[8][3]^2[2]^2}{[6]} u_{1}^{9}  &  \sqrt{\frac{[10][3]^3}{[6]^3[5]}} u_{1}^{10}  &  \sqrt{\frac{[10][9][8][4]}{[7][5]^2[3][2]}}\frac{[3]^2}{[6]^2} u_{1}^{11}  &  \sqrt{\frac{[10][9][8]}{[7][6][5][4]}}\frac{[3]}{[6]} u_{1}^{12} \\
\\
\rule[-.3\baselineskip]{0pt}{6ex} \sqrt{\frac{[10][4]}{[7][5]}}\frac{[4][3]^2}{[6]^2[2]} u_{2}^{1}  &  \sqrt{\frac{[2]}{[6][3]}}\frac{1}{[8][5]}u_{2}^{2}  &  \sqrt{[10][7][4]}\frac{[3]}{[6]^2}u_{2}^{3}  &  \frac{1}{[8][5]\sqrt{[3]}}u_{2}^{4}  &  \sqrt{\frac{[10][7][4][3]}{[6]^2[5][2]^2}}u_{2}^{5}  &  \sqrt{\frac{[10][4]}{[5]}}\frac{[4]}{[8][6][2]}u_{2}^{6}  &  \sqrt{\frac{1}{[7][5]}}\frac{[3][2]}{[8][6]}u_{2}^{7}  &  \sqrt{\frac{[9][7][4]!}{[8]^2[5]}}\frac{[4][3]}{[6]^2}u_{2}^{8}  &  \sqrt{\frac{[7][3][2]}{[6]}}\frac{[2]}{[8][6]}u_{2}^{9}  &  \sqrt{\frac{[7][4]^5[3]}{[8]^2[6]^3}}\frac{[3]}{[5][2]}u_{2}^{10}  &  \sqrt{\frac{[9][2]^3}{[8][5][3]}}\frac{[3]^2}{[6]^2}u_{2}^{11}  &  \sqrt{\frac{[9]}{[8][6]}}\frac{[3][2]}{[6][5]}u_{2}^{12} \\
\\
\rule[-.3\baselineskip]{0pt}{6ex} \sqrt{\frac{[3]^3[2]}{[6]^3[5]}} u_{3}^{1}  &  \sqrt{\frac{[10][7]}{[4]}}\frac{[5]}{[6][3]}u_{3}^{2}  &  \sqrt{\frac{[2]}{[6]^3[3]}}u_{3}^{3}  &  \sqrt{\frac{[10][7]}{[6][4][3]^2[2]}}u_{3}^{4}  &  \sqrt{\frac{1}{[6][5][3]^2[2]}}u_{3}^{5}  &  \sqrt{\frac{[7]^3[5]}{[6][3]^3[2]}}u_{3}^{6}  &  \sqrt{\frac{[10]}{[6]!}}u_{3}^{7}  &  \sqrt{\frac{[10]^3[9]}{[6][5]}}\frac{[7][3]}{[6][2]}u_{3}^{8}  &  \sqrt{\frac{[10]}{[4]}}\frac{1}{[6][3]}u_{3}^{9}  &  \sqrt{\frac{[10]}{[2]}}\frac{[7]}{[3][2]}u_{3}^{10}  &  \sqrt{\frac{[10][9][8][7]}{[6][5][4]}}\frac{[3]}{[6]}u_{3}^{11}  &  0 \\
\\
\rule[-.3\baselineskip]{0pt}{6ex} \sqrt{\frac{[10][4]}{[7][5]}}\frac{[3]^2}{[6][2]} u_{4}^{1}  &  \sqrt{\frac{[3][2]}{[6]}}\frac{[2]^2}{[8][5]}u_{4}^{2}  &  \sqrt{\frac{[10][7][4]}{[2]^2}}\frac{[3]}{[6]}u_{4}^{3}  & \frac{[2]^2}{[8][7][5]\sqrt{[3]}}u_{4}^{4}  &  \sqrt{\frac{[10][7][4]}{[8]^2[5][3]}}\frac{[4]}{[2]^2}u_{4}^{5}  &  \sqrt{\frac{[10][4]}{[8]^2[5]}}\frac{[4]}{[7][2]^2}u_{4}^{6}  &  \frac{[2]^2}{[8]\sqrt{[7][5]}}u_{4}^{7}  &  \sqrt{\frac{[9][4][3]}{[7][5][2]}}\frac{[4][3]}{[8][6]}u_{4}^{8}  &  \sqrt{\frac{[7][3][2]}{[6]}}\frac{[2]^2}{[8]}u_{4}^{9}  &  \sqrt{\frac{[4][3]}{[7][6]}}\frac{[4]}{[8][5][2]}u_{4}^{10}  &  \sqrt{\frac{[9][3]}{[8][5][2]}}\frac{[3][2]}{[7][6]}u_{4}^{11}  &  \sqrt{\frac{[9]}{[8][6]}}\frac{[3]^2}{[7][5]}u_{4}^{12} \\
\\
\rule[-.3\baselineskip]{0pt}{6ex} \frac{[14][3]^2[2]}{[7][6][5]} u_{5}^{1}  &  \sqrt{\frac{[10][7]}{[6]!}}\frac{[9][4]}{[8][3]}u_{5}^{2}  &  \frac{[2]^2}{[6]\sqrt{[5]}}u_{5}^{3}  &  \sqrt{\frac{[10][7]}{[5]![2]}}\frac{[4]}{[8]}u_{5}^{4}  &  \frac{[4][2]}{[8][5]\sqrt{[3]}}u_{5}^{5}  &  \frac{[4][2]\sqrt{[7]}}{[8][3]}u_{5}^{6}  &  \frac{\sqrt{[10][4]}}{[8][5][2]}u_{5}^{7}  &  \sqrt{\frac{[10]![3]}{[8]![2]^3}}\frac{[7][5]!}{[8][6]}u_{5}^{8}  &  \sqrt{\frac{[10][4]^2}{[8]^2[6]!}}u_{5}^{9}  &  \sqrt{\frac{[10][3]}{[6][5]}}\frac{[7][4]}{[8][2]}u_{5}^{10}  &  \sqrt{\frac{[10][9][7][4][3]^3}{[8][6]^2[5]^2[2]}}u_{5}^{11}  &  \sqrt{\frac{[10]![4]}{[8]^2[6][5]}}\frac{[3]^2}{[2]}u_{5}^{12} \\
\\
\rule[-.3\baselineskip]{0pt}{6ex} \frac{[3]}{\sqrt{[7]}[6]} u_{6}^{1}  &  \sqrt{\frac{[10][4][2]}{[6][5][3]^3}}\frac{[4]}{[8]}u_{6}^{2}  &  \frac{\sqrt{[7][5]}}{[6][3]}u_{6}^{3}  &  \sqrt{\frac{[10][4]}{[5][3]^3}}\frac{[4]}{[8][7]}u_{6}^{4}  &  \sqrt{\frac{[7]}{[3]}}\frac{[4]}{[8][3][2]}u_{6}^{5}  &  \frac{[4]}{[8][7][3]^2[2]}u_{6}^{6}  &  \sqrt{\frac{[10][4]}{[7]}}\frac{[4]}{[8][3]}u_{6}^{7}  &  \sqrt{\frac{[10][9][3]}{[7][6]^2[2]}}\frac{[4]}{[8]}u_{6}^{8}  & \sqrt{\frac{[10][7]!}{[8]^2}}\frac{[4][2]^2}{[6][3]^2}u_{6}^{9}  &  \sqrt{\frac{[10][4]}{[7]![2]}}\frac{[4]}{[8]}u_{6}^{10}  &  \sqrt{\frac{[10][9][4]!}{[8][7]^2}}\frac{[4]}{[6][2]^2}u_{6}^{11}  &  \sqrt{\frac{[10][9][4]}{[8][6][5]}}\frac{[3]}{[7][2]}u_{6}^{12} \\
\\
\rule[-.3\baselineskip]{0pt}{6ex} \sqrt{\frac{[10][4]}{[5]^2}}\frac{[3]!}{[7][6]} u_{7}^{1}  &  \sqrt{\frac{[4]}{[7]!}}\frac{[2]}{[8]}u_{7}^{2}  &  \sqrt{\frac{[10][4]}{[5]}}\frac{[12][2]}{[6]^2}u_{7}^{3}  &  \frac{1}{\sqrt{[8]^2[7][5][3]}}u_{7}^{4}  & \sqrt{\frac{[10][4]}{[3]}}\frac{[4][2]}{[5]}u_{7}^{5}  &  \sqrt{\frac{[10][4]}{[7]}}\frac{[4]}{[8][3]}u_{7}^{6}  &  \frac{1}{[8][7][5]}u_{7}^{7}  &  \sqrt{\frac{[9]}{[4]!}}\frac{[4]!}{[8][6]}u_{7}^{8}  &  \sqrt{\frac{[2]}{[6][5][3]}}\frac{1}{[8]}u_{7}^{9}  &  \sqrt{\frac{[4][3]}{[6][5]}}\frac{[12][4]}{[8][6][2]}u_{7}^{10}  &  \sqrt{\frac{[9][3][2]}{[8][7]}}\frac{[18][3]}{[9][6]^2[5]}u_{7}^{11}  &  \sqrt{\frac{[9]}{[8][7][6][5]}}[3]u_{7}^{12} \\
\\
\rule[-.3\baselineskip]{0pt}{6ex} \sqrt{[10][9][3]!}\frac{[5][3]}{[6]^2} u_{8}^{1}  &  \sqrt{\frac{[9][7][4]}{[6][5][2]^2}}\frac{[4]}{[8]}u_{8}^{2}  &  \sqrt{\frac{[10]![5]!}{[8]![4]}}\frac{[7][3]}{[6]^2}u_{8}^{3}  &  \sqrt{\frac{[9][4]}{[7][5][2]^3}}\frac{[4]}{[8]}u_{8}^{4}  &  \sqrt{\frac{[10][9]}{[2]^3}}\frac{[7]!}{[8][6]^2}u_{8}^{5}  &  \sqrt{\frac{[10][9][3]}{[7][2]}}\frac{[4]}{[8][6]}u_{8}^{6}  &  \frac{\sqrt{[9][4]!}}{[8][6]}u_{8}^{7}  &  \frac{[4]![3]}{[8][6]^2}u_{8}^{8}  &  \sqrt{\frac{[9][5][4]}{[6]}}\frac{[7][3]!}{[8][6]}u_{8}^{9}  &  \sqrt{\frac{[9][2]}{[6][5]}}\frac{[4][3]}{[8][6]}u_{8}^{10}  &  \sqrt{\frac{[8]}{[7][4]}}\frac{[3]^2[2]}{[6]^2}u_{8}^{11}  &  \sqrt{\frac{[4][3]^2[2]}{[8][7][6]^3[5]}}u_{8}^{12} \\
\\
\rule[-.3\baselineskip]{0pt}{6ex} \sqrt{\frac{[10][4]!}{[6][5]}}\frac{[8][3]!}{[6][5]} u_{9}^{1}  &  \frac{\sqrt{[7]}[2]}{[8][6][3]}u_{9}^{2}  &  \sqrt{\frac{[10][4][2]}{[6][3]}}\frac{[2]}{[6]}u_{9}^{3}  &  \sqrt{\frac{[7]}{[6][2]}}\frac{[2]}{[8][3]}u_{9}^{4}  &  \sqrt{\frac{[10][4][2]}{[6][5][3]^2}}\frac{[4]}{[8]}u_{9}^{5}  &  \sqrt{\frac{[10]^3[7][4]!}{[8]^2[6][5]}}\frac{[4]!}{[3]^3}u_{9}^{6}  &  \sqrt{\frac{[2]}{[6][5][3]}}\frac{1}{[8]}u_{9}^{7}  &  \sqrt{\frac{[9][5]}{[6][4]}}\frac{[7][4]!}{[8][6]}u_{9}^{8}  &  \frac{[2]}{[8][6][3]}u_{9}^{9}  &  \sqrt{\frac{[4]}{[2]}}\frac{[7][4]}{[8][6]}u_{9}^{10}  &  \sqrt{\frac{[9][7]}{[8][6][5]}}\frac{[18][3]^2[2]}{[9][6]^2}u_{9}^{11}  &  \sqrt{\frac{[9][7][3]!}{[8]}}\frac{[3]}{[6]}u_{9}^{12} \\
\\
\rule[-.3\baselineskip]{0pt}{6ex} \sqrt{\frac{[10][3]}{[6][5]}}\frac{[3]}{[6]} u_{10}^{1} &  \frac{[7][4]}{[2]}\frac{[4]}{[8][6][5]}u_{10}^{2}  &  \sqrt{\frac{[10][3]}{[6]}}\frac{[7]}{[6]}u_{10}^{3}  &  \sqrt{\frac{[4][3]^2}{[7][6]}}\frac{[4]^2}{[8][5]!}u_{10}^{4}  &  \sqrt{\frac{[10]}{[6][5]}}\frac{[7][4]}{[8][2]}u_{10}^{5}  &  \sqrt{\frac{[10][4]}{[7]![2]}}\frac{[4]}{[8]}u_{10}^{6}  &  \sqrt{\frac{[4][3]}{[6][5]}}\frac{[12][4]}{[8][6][2]}u_{10}^{7}  &  \sqrt{\frac{[9][4]!}{[6]![2]}}\frac{[4]!}{[8][6]}u_{10}^{8}  &  \sqrt{\frac{[4]}{[2]}}\frac{[7][4]}{[8][6]}u_{10}^{9}  &  \frac{[4][3]}{[8][6][5]}u_{10}^{10}  &  \sqrt{\frac{[9][4]^3[3]}{[8]!}}\frac{[3]}{[6]}u_{10}^{11}  &  \sqrt{\frac{[9][4][3]}{[8][7]}}\frac{[3]}{[6][5]}u_{10}^{12} \\
\\
\rule[-.3\baselineskip]{0pt}{6ex} \sqrt{\frac{[10]![4]}{[7]![7][3]!}}\frac{[3]^2}{[6]^2[5]} u_{11}^{1}  &  \sqrt{\frac{[9]}{[8][6][5]}}[3]u_{11}^{2}  &  \sqrt{\frac{[10]![4]^2}{[6]![5]!}}\frac{[3]^2}{[6]^2}u_{11}^{3}  &  \sqrt{\frac{[9]}{[8][5][2]}}\frac{[3]}{[7]}u_{11}^{4}  &  \sqrt{\frac{[10]![4]^5}{[6]!^2[2]}}\frac{[3]^3}{[8]}u_{11}^{5}  &  \sqrt{\frac{[10][9][4]!}{[8][7]^2[2]^4}}\frac{[4]}{[6]}u_{11}^{6}  &  \sqrt{\frac{[6]![3]!}{[9]!}}\frac{[18][3]}{[6]^2[5]}u_{11}^{7}  &  \sqrt{\frac{[8]}{[7][4]}}\frac{[3]^2[2]}{[6]^2}u_{11}^{8}  &  \sqrt{\frac{[4]!}{[9]!}}\frac{[18][7][3]!^2}{[6]^2[2]}u_{11}^{9}  &  \sqrt{\frac{[9][3]}{[8]!}}\frac{[4]^2[3]}{[6]}u_{11}^{10}  &  \frac{[8][3]^2}{[7][6]^2[5]}u_{11}^{11}  &  \sqrt{\frac{[3]}{[6][5][2]}}\frac{[4][3]}{[7][6]}u_{11}^{12} \\
\\
\rule[-.3\baselineskip]{0pt}{6ex} \sqrt{\frac{[10]![3]!}{[7]!^2}}\frac{[3]}{[6]} u_{12}^{1}  &  \sqrt{\frac{[9][3]!}{[8]}}\frac{1}{[6][5]}u_{12}^{2}  &  \sqrt{\frac{[10]!}{[6]![6][4]}}\frac{[3]}{[6]}u_{12}^{3}  &  \sqrt{\frac{[9][3]}{[8][6]}}\frac{[3]}{[7][5]}u_{12}^{4}  &  \sqrt{\frac{[10]![4]!^2}{[6]!^2[2]^3}}\frac{[3]}{[8]}u_{12}^{5}  &  \sqrt{\frac{[10][9][4]}{[8][6][5]}}\frac{[3]}{[7][2]}u_{12}^{6}  &  \sqrt{\frac{[9][4]!}{[8]!}}[3]u_{12}^{7}  &  \sqrt{\frac{[4][3]^3[2]}{[8][7][6]^3[5]}}u_{12}^{8}  &  \sqrt{\frac{[9][7][3]!}{[8]}}\frac{[3]}{[6]}u_{12}^{9}  &  \sqrt{\frac{[9][4][3]}{[8][7]}}\frac{[3]}{[6][5]}u_{12}^{10}  &  \sqrt{\frac{[3]}{[6][5][2]}}\frac{[4][3]}{[7][6]}u_{12}^{11}  &  \frac{[3]^2}{[7][6][5]}u_{12}^{12}
\end{array}\hspace{-0.5mm}\right)\nn
$}}
\end{landscape}

\section{Exclusive matrices from inclusive ones}
\label{trick}
In the paper \cite{MMMStrick} it was described a method for  multiplicity free cases only how to calculate exclusive quantum Racah matrices $S$ and $\bar S$ if inclusive matrices are known. Since $R=[3,3]$ is exactly the case, then we can directly apply that method to our matrices.

The idea is based on the fact that there is a two-parametric family, which is simultaneously 3-strand braid $(m,-1|\pm n,-1)$ and pretzel $Pr(m,n,\pm\bar 2)$. From one hand, for this 3-strand braid family formula (\ref{3str}) reduces to
\be
H_R^{(m,-1|\pm n,-1)} = \sum_{Y,Z\in R^{\otimes 2}} h_{YZ}\cdot \lambda_Y^m \lambda_Z^n
\label{33str}
\ee
which is the usual evolution formula and where  $\lambda_Y$ is the eigenvalue (\ref{evY}). From another hand, for this pretzel family formula (\ref{arbor}) simplifies to:
\be
H_R^{Pr(m,n,\overline{\pm 2})}   =   d_R\sum_{\bar X\in R\otimes \bar R} \frac{(ST^mS^\dagger)_{\emptyset \bar X}  (ST^nS^\dagger)_{\emptyset \bar X} (\bar S \bar T^{\pm 2} \bar S)_{\emptyset \bar X}} {S_{\emptyset \bar X}}   =    d_R^{-1} \!\!\!\! \sum_{\stackrel{\bar X\in R\otimes \bar R}{ Y,Z\in R\otimes R}} \sqrt{ d_Yd_Z} K_{\bar X} S_{\bar X Y}S_{\bar X Z}\cdot \lambda_Y^m\lambda_Z^n
\label{Hpre}
\ee
where the square $S^2_{\emptyset Y} =d_Y/d_R^2$ and
$
K_{\bar X} = d_Rd_X^{-1/2}(\bar S \bar T^{\pm 2} \bar S)_{\emptyset \bar X}
\label{KX}
$. The pretzel formula (\ref{Hpre}) should be compared with the answer (\ref{33str}). It gives:
\be\label{12}
\sum_X  K_{\bar X} S_{\bar X Y} S_{\bar XZ} = \frac{d_R}{\sqrt{d_Y d_Z}}\cdot h_{YZ}    =    \mathfrak{h}_{YZ}
\ee
i.e. $F_{\bar X}$ are the eigenvalues of the matrix $\mathfrak{h}$ at the r.h.s., while our needed $S_{\bar XY}$ is the orthogonal diagonalizing matrix (made from the normalized eigenvectors). It provides the explicit way to calculate $S$. Then we can extract $\bar S$ just from the known $S$ using relation (63) from \cite{MMMRS}:
\be
\bar S =   \bar T^{-1} S T^{-1} S^\dagger \bar T^{-1}
\label{bSfromS}
\ee

\bigskip

Our calculations of matrices $S$ and $\bar S$ confirms formulas obtained by A.Morozov in the paper \cite{Morozov33}.




\section{Examples of $[3,3]$-colored HOMFLY
for knots, which are 3-strand  but not arborescent}

Using the manifest expressions for the inclusive Racah matrices, one can evaluate the HOMFLY polynomials of all 3-strand knots in representation $[3,3]$. The results for arborescent knots are in \cite{knotebook} due to results of \cite{Morozov33}, for different 3-strand knots can be found in \cite{knotebook}, here, as an illustration, we write down the answers for two knots $10_{100}$ (in a table style) and $10_{161}$ (in a polynomial style) that are 3-strand and can not be presented by arborescent diagrams:

\begin{figure}[h!]
\centering\leavevmode
\includegraphics[width=6 cm]{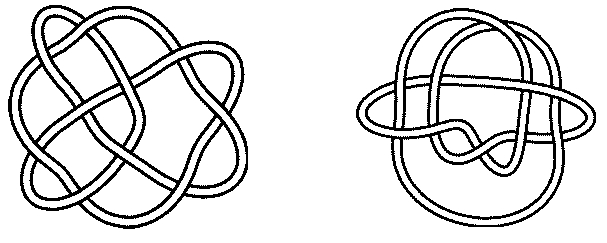}
\caption{Knots $10_{100}$ and $10_{161}$ from the Katlas \cite{katlas}}
\label{knots}
\end{figure}

\bigskip

{\tiny
\vspace*{-1.5cm}
\be
{\normalsize \text{\bf Knot $10_{100}$}:}
\begin {array}{c|ccccccccccccc}
q\backslash A &{A}^{12}&{A}^{14}&{A}^{16}&{A}^{18}&{A}^{20}&{A}^{22}&{A}^{24}&{A}^{26}
&{A}^{28}&{A}^{30}&{A}^{32}&{A}^{34}&{A}^{36}
\\ \hline {-108}&0&0&0&0&0&-1&1&0&0&0&0&0&0   \\ {-106}&0&0&0&0&2&0&-2&0&0&0&0&0&0   \\ {-104}&0&0&0&-1&-1&5&-2&-1&0&0&0&0&0   \\ {-102}&0&0&0&-2&-7&3&6&0&0&0&0&0&0   \\ {-100}&0&0&2&6&-8&-13&8&5&0&0&0&0&0   \\ {-98}&0&0&-1&14&18&-26&-10&3&2&0&0&0&0   \\ {-96}&0&-1&-7&-6&50&18&-38&-14&-1&0&0&0&0   \\ {-94}&0&0&-10&-50&-4&90&7&-26&-7&0&0&0&0   \\ {-92}&0&5&14&-52&-128&37&108&20&-7&-1&0&0&0   \\ {-90}&1&6&55&81&-150&-183&71&94&19&-2&0&0&0   \\ {-88}&-2&-9&24&224&153&-292&-184&34&48&6&0&0&0   \\ {-86}&-3&-35&-102&77&526&130&-365&-195&-11&13   &0&0&0\\ {-84}&3&-15&-207&-410&245&775&88&-295&-   132&-8&2&0&0\\ {-82}&11&71&-5&-684&-797&530&878&   154&-133&-49&-1&0&0\\ {-80}&8&123&453&61&-1441&-   1015&696&805&184&-40&-7&0&0\\ {-78}&-26&-10&548&   1408&-5&-2192&-1030&500&522&95&-8&0&0\\ {-76}&-   37&-282&-219&1519&2732&-367&-2494&-1105&155&209&16&-1&0   \\ {-74}&8&-299&-1345&-850&3054&3654&-542&-2205&   -901&12&50&0&0\\ {-72}&82&186&-1101&-3743&-1389&   4565&4052&-179&-1361&-457&6&5&0\\ {-70}&82&755&   1104&-2702&-6887&-1524&5248&3893&324&-582&-115&5&0   \\ {-68}&-75&516&3115&3173&-5065&-9570&-1498&   4422&2966&313&-169&-11&0\\ {-66}&-198&-717&1689&   7889&5707&-7632&-10646&-2159&2637&1519&82&-32&1   \\ {-64}&-103&-1590&-3181&3581&14265&7248&-8482&-10017&-2379&1118&486&-   2&-2\\ {-62}&216&-595&-5921&-8349&6613&19672&   8333&-6909&-7305&-1588&375&83&-3\\ {-60}&386&   1786&-1594&-14201&-14562&9523&22263&8938&-3721&-3891&-531&98&4   \\ {-58}&44&2714&6999&-3033&-24753&-19828&10562&   20355&8097&-1465&-1396&-78&11\\ {-56}&-470&209&   9581&16991&-5053&-34559&-22683&7431&14738&5015&-536&-314&4   \\ {-54}&-560&-3517&74&21684&29662&-7720&-38581&   -23416&3103&7958&1953&-180&-32\\ {-52}&119&-3818   &-12632&-952&37605&40340&-7184&-35296&-19569&518&3114&421&-33   \\ {-50}&846&885&-12847&-29533&-1485&51537&47263   &-2882&-25030&-12198&155&809&33\\ {-48}&620&5721   &3648&-28625&-50191&-2206&58300&47015&2458&-13816&-5108&123&100   \\ {-46}&-423&4344&19688&8994&-48381&-69694&-   4661&51984&38762&3530&-5689&-1322&50\\ {-44}&-   1206&-2842&14436&43985&15508&-67257&-80760&-11413&36961&24129&1865&-   1618&-158\\ {-42}&-540&-7881&-9599&31546&74801&   20877&-74691&-80645&-15445&20216&10728&434&-232\\ {-40}&895&-3901&-26156&-21813&53953&102763&28769&-66749&-65105&-   13111&8668&3062&18\\ {-38}&1390&5416&-13036&-   57593&-35632&74203&121056&36206&-45855&-41325&-6650&2562&415   \\ {-36}&273&9188&17389&-28989&-96403&-49911&   83363&118882&38455&-25066&-19014&-1998&417\\ {-   34}&-1364&2226&30038&37032&-49719&-134233&-62360&71918&96887&29022&-   10510&-5805&-269\\ {-32}&-1360&-7953&8304&65359&   60633&-70254&-157152&-72805&48566&61497&15141&-3146&-864   \\ {-30}&235&-9091&-24680&19432&109898&82886&-   77057&-156109&-69312&24660&29375&4918&-513\\ {-   28}&1595&470&-29687&-51387&35954&152984&103522&-65824&-125924&-51700&   9810&9251&755\\ {-26}&1096&9529&-557&-64745&-   82739&51642&181807&113317&-40054&-81468&-27120&2678&1485   \\ {-24}&-806&7318&28957&-4749&-109410&-113888&   57585&179495&106291&-18074&-39162&-9319&389\\ {-   22}&-1564&-3388&24549&59685&-13547&-154685&-139416&44103&147012&77070&   -5118&-12653&-1524\\ {-20}&-518&-9488&-7924&   54599&95416&-23486&-184505&-152084&22693&94460&41669&-760&-2065   \\ {-18}&1120&-4106&-28689&-11136&94217&131413&-   24850&-185841&-138071&3300&46102&14571&0\\ {-16}   &1304&5565&-15353&-58410&-10881&136018&161669&-14432&-150912&-101625&-   3164&14841&2541\\ {-14}&-210&7643&13896&-36333&-   93932&-7537&166506&173457&4960&-98743&-54777&-2826&2386   \\ {-12}&-1155&453&23236&23271&-67070&-129533&-   9218&168327&159792&14943&-47242&-19656&-678\\ {-   10}&-703&-6175&4284&47863&28004&-100771&-160159&-19838&140854&115826&   14945&-15138&-3501\\ {-8}&656&-4281&-15829&15413   &76543&30778&-128190&-175659&-30233&89705&63823&7299&-2346   \\ {-6}&1010&2616&-13883&-26391&33341&108183&   34482&-135742&-159730&-35565&43198&22841&1645\\    {-4}&-89&5044&4481&-28739&-34245&58881&136953&40980&-112069&-119525&-   25672&13002&4054\\ {-2}&-798&528&12883&3071&-   48444&-37070&81730&151678&50800&-73594&-64698&-11892&1969   \\ 0&-530&-4079&2502&21733&-6282&-69554&-40804&90488   &145516&46586&-32563&-23509&-2585\\ {2}&566&-   2467&-9113&7462&26006&-21322&-93713&-48982&81011&106510&33550&-9274&-   4071\\ {4}&784&2603&-6351&-12281&13561&28028&-   40297&-113646&-51360&48833&59834&14483&-1135\\ {   6}&-171&3703&6358&-8896&-10447&25571&30479&-55407&-109351&-51045&21122   &21015&3074\\ {8}&-713&-550&8772&10760&-8525&58&   43864&33924&-47752&-86938&-33327&4392&3714\\ {10   }&-425&-4099&-2361&13285&13501&-4248&17228&61290&44883&-32217&-46545&-   14951&150\\ {12}&585&-1826&-10674&-6503&12539&   11602&-2257&28325&73474&39809&-9951&-16684&-2969\\ {14}&714&3061&-4415&-19530&-15116&4600&615&-11823&34545&55952&29844   &-1168&-2789\\ {16}&-174&3743&9198&-5696&-28126&   -25472&-8060&-24086&-17189&17410&32686&12468&459\\ {18}&-691&-673&10272&19086&-3040&-32224&-28222&-26231&-33097&-26867   &6400&10671&2646\\ {20}&-402&-4220&-2412&19326&   33379&4937&-21082&-25836&-22612&-35731&-18175&-780&1615   \\ {22}&539&-2044&-11957&-6355&28929&46817&21311   &-6737&-4284&-18428&-18017&-8745&-505\\ {24}&671   &2780&-5916&-24534&-13140&32229&57240&27327&17957&1929&-3695&-6427&-   1794\\ {26}&-101&3785&8315&-11617&-39692&-26295&   29715&43424&34831&15514&7678&70&-756\\ {28}&-630   &-216&11075&16417&-16644&-57260&-35891&6363&32826&18062&11867&3350&500   \\ {30}&-401&-3801&-343&22087&28823&-20948&-   60265&-50758&-2047&9304&9243&3052&696\\ {32}&413   &-2287&-10760&-721&36871&40349&-10946&-60175&-39022&-13665&4084&1358&   467\\ {34}&597&2048&-6920&-21811&-28&47825&57941   &-5391&-33543&-31867&-6006&53&-35\\ {36}&-3&3471   &5874&-14595&-34021&-4872&59041&54899&14229&-22051&-13059&-2458&131   \\ {38}&-499&357&10251&10617&-22783&-50711&-5981   &45286&55984&10085&-6702&-4556&-191\\ {40}&-364&   -2943&1556&20244&17701&-33913&-56173&-21890&38439&32921&8980&-2584&-   681\\ {42}&255&-2246&-8329&3488&33465&21604&-   32364&-65849&-18289&16322&19260&2595&-343\\ {44}   &458&1160&-6951&-16618&7588&42853&33659&-35586&-48289&-22020&9018&5979   &379\\ {46}&53&2726&3073&-14724&-24829&7432&   56145&30252&-17584&-37940&-10872&2281&1169\\ {48   }&-327&716&8163&5013&-22772&-36407&12830&49020&36918&-13287&-17634&-   4663&469\\ {50}&-270&-1865&2740&16264&8402&-   33487&-37957&3475&47668&23594&-3474&-6705&-832\\ {52}&123&-1791&-5209&6194&26757&8597&-33790&-46073&6466&27988&15744&-   1270&-1103\\ {54}&287&432&-5746&-10242&11545&   33469&15285&-39578&-34322&-2204&16162&5522&-182\\ {56}&58&1731&756&-12424&-15152&13055&43053&9814&-26981&-28924&-960&   5193&1135\\ {58}&-170&701&5212&348&-19562&-22693   &18633&37090&14912&-21332&-14609&-1251&950\\ {60   }&-155&-920&2697&10469&654&-28265&-22435&13037&36280&8524&-9256&-5829&   -303\\ {62}&46&-1096&-2408&6333&17500&-716&-   28576&-26671&15067&22458&7000&-3283&-1032\\ {64}   &139&59&-3625&-4419&11576&22146&3017&-31896&-18285&6856&13183&2710&-   519\\ {66}&30&840&-322&-8030&-6241&14033&28205&-   978&-22715&-15570&3766&4488&634\\ {68}&-70&420&   2519&-1700&-13011&-10155&18017&23603&2405&-17387&-8032&680&828   \\ {70}&-59&-346&1711&4995&-2986&-18998&-9792&   14201&21925&337&-8049&-3365&27\\ {72}&19&-472&-   760&4205&8507&-4935&-19561&-11898&13993&13336&1677&-2810&-635   \\ {74}&47&-12&-1636&-1110&7847&11035&-3330&-   20712&-7211&7602&7702&855&-441\\ {76}&2&291&-369   &-3761&-1211&10034&14325&-5106&-14623&-6239&3856&2667&260   \\ {78}&-21&139&870&-1464&-6310&-2606&12422&   11748&-2639&-10454&-3311&934&494\\ {80}&-9&-101&   683&1675&-2691&-9502&-2319&10248&10294&-2260&-4797&-1481&87   \\ {82}&7&-122&-153&1842&2966&-4141&-10036&-3282   &9065&6030&-305&-1601&-295\\ {84}&7&3&-484&-49&   3627&4018&-3691&-10225&-1436&4995&3404&116&-239\\ {86}&-3&61&-140&-1215&219&4884&5433&-4248&-7009&-1504&2300&1173&84   \\ {88}&-2&18&199&-635&-2187&-133&5992&4315&-   2771&-4631&-923&546&214\\ {90}&1&-18&158&375&-   1240&-3502&-2&5024&3509&-1838&-2030&-488&43\\ {   92}&0&-13&-17&525&735&-1910&-3811&-332&4085&1927&-468&-619&-106   \\ {94}&0&2&-85&57&1141&1119&-1830&-3732&203&   2150&1071&-17&-79\\ {96}&0&5&-25&-265&193&1643&   1622&-1939&-2413&-71&863&359&29\\ {98}&0&0&28&-   158&-542&102&2016&1220&-1307&-1432&-155&167&61\\ {100}&0&-1&18&64&-335&-963&111&1682&866&-758&-565&-120&1   \\ {102}&0&0&-2&92&149&-522&-1096&45&1255&421&-   175&-141&-26\\ {104}&0&0&-7&8&231&279&-498&-1011   &183&597&227&0&-9\\ {106}&0&0&-1&-37&35&363&430&   -516&-594&59&187&66&8\\ {108}&0&0&2&-17&-94&-4&   444&303&-345&-301&-16&21&7\\ {110}&0&0&0&10&-43&   -188&-27&363&176&-175&-94&-19&-3\\ {112}&0&0&0&6   &27&-66&-223&-17&244&72&-27&-14&-2\\ {114}&0&0&0   &-2&22&60&-50&-190&28&99&30&2&1\\ {116}&0&0&0&-1   &-1&38&94&-62&-98&6&19&5&0\\ {118}&0&0&0&0&-7&-   13&42&60&-41&-39&-2&0&0\\ {120}&0&0&0&0&-1&-17&-   22&37&28&-17&-7&-1&0\\ {122}&0&0&0&0&2&0&-20&-13   &22&10&-1&0&0\\ {124}&0&0&0&0&0&5&5&-17&-1&6&2&0   &0\\ {126}&0&0&0&0&0&0&9&0&-7&-2&0&0&0   \\ {128}&0&0&0&0&0&-1&-2&5&-1&-1&0&0&0   \\ {130}&0&0&0&0&0&0&-2&0&2&0&0&0&0   \\ {132}&0&0&0&0&0&0&1&-1&0&0&0&0&0
\end {array}
\ee
}

\paragraph{Knot $10_{161}$:}

{\footnotesize
$H_{[3,3]}^{10_{161}} = \Big( {q}^{78}-{q}^{74}-2\,{q}^{72}+3\,{q}^{68}+2\,{q}^{66}+{q}^{64}-3\,{q}^{62}-2\,{q}^{60}+4\,{q}^{54}+{q}^{50}-2\,{q}^{48}-4\,{q}^{46}+{q}^{44}+{q}^{42}+5\,{q}^{40}-4\,{q}^{36}-3\,{q}^{34}-2\,{q}^{32}+4\,{q}^{30}+3\,{q}^{28}+{q}^{26}-2\,{q}^{24}-3\,{q}^{22}+{q}^{18}+{q}^{16} \Big) {A}^{60}+ \Big( -{q}^{84}-{q}^{82}+{q}^{80}+4\,{q}^{78}+5\,{q}^{76}-{q}^{74}-8\,{q}^{72}-9\,{q}^{70}-{q}^{68}+10\,{q}^{66}+11\,{q}^{64}+2\,{q}^{62}-9\,{q}^{60}-13\,{q}^{58}-7\,{q}^{56}+4\,{q}^{54}+12\,{q}^{52}+13\,{q}^{50}+2\,{q}^{48}-13\,{q}^{46}-17\,{q}^{44}-5\,{q}^{42}+17\,{q}^{40}+23\,{q}^{38}+8\,{q}^{36}-14\,{q}^{34}-22\,{q}^{32}-7\,{q}^{30}+13\,{q}^{28}+21\,{q}^{26}+9\,{q}^{24}-8\,{q}^{22}-14\,{q}^{20}-7\,{q}^{18}+4\,{q}^{16}+7\,{q}^{14}+4\,{q}^{12}-{q}^{10}-3\,{q}^{8}-2\,{q}^{6}-{q}^{4} \Big) {A}^{58}+ \Big( 2+5\,{q}^{-2}+2\,{q}^{-6}+3\,{q}^{-4}-39\,{q}^{44}-55\,{q}^{42}-15\,{q}^{40}+33\,{q}^{38}+57\,{q}^{36}+20\,{q}^{34}-38\,{q}^{32}-61\,{q}^{30}-33\,{q}^{28}+24\,{q}^{26}+49\,{q}^{24}+29\,{q}^{22}-16\,{q}^{20}-38\,{q}^{18}-17\,{q}^{58}-35\,{q}^{56}-20\,{q}^{54}+8\,{q}^{52}+43\,{q}^{50}+37\,{q}^{48}+5\,{q}^{46}-2\,{q}^{72}-19\,{q}^{70}-21\,{q}^{68}-4\,{q}^{66}+27\,{q}^{64}+32\,{q}^{62}+17\,{q}^{60}-{q}^{86}-4\,{q}^{84}-8\,{q}^{82}-7\,{q}^{80}+3\,{q}^{78}+11\,{q}^{76}+14\,{q}^{74}+{q}^{88}-26\,{q}^{16}+4\,{q}^{14}+23\,{q}^{12}+18\,{q}^{10}+5\,{q}^{8}-7\,{q}^{4}-7\,{q}^{6}-2\,{q}^{2} \Big) {A}^{56}+ \Big( -16-{q}^{-2}-{q}^{-16}-3\,{q}^{-14}-4\,{q}^{-12}-4\,{q}^{-10}+4\,{q}^{-6}+5\,{q}^{-4}-3\,{q}^{44}-77\,{q}^{42}-90\,{q}^{40}-26\,{q}^{38}+72\,{q}^{36}+104\,{q}^{34}+49\,{q}^{32}-48\,{q}^{30}-96\,{q}^{28}-46\,{q}^{26}+43\,{q}^{24}+97\,{q}^{22}+64\,{q}^{20}-14\,{q}^{18}+13\,{q}^{58}-48\,{q}^{56}-83\,{q}^{54}-61\,{q}^{52}+8\,{q}^{50}+65\,{q}^{48}+65\,{q}^{46}+24\,{q}^{72}-5\,{q}^{70}-38\,{q}^{68}-48\,{q}^{66}-16\,{q}^{64}+30\,{q}^{62}+47\,{q}^{60}+11\,{q}^{86}+8\,{q}^{84}-2\,{q}^{82}-11\,{q}^{80}-7\,{q}^{78}+10\,{q}^{76}+27\,{q}^{74}+{q}^{92}+4\,{q}^{90}+8\,{q}^{88}-64\,{q}^{16}-53\,{q}^{14}+{q}^{12}+37\,{q}^{10}+36\,{q}^{8}-20\,{q}^{4}+7\,{q}^{6}-25\,{q}^{2} \Big) {A}^{54}+ \Big( -33-17\,{q}^{-2}-2\,{q}^{-14}-3\,{q}^{-12}+5\,{q}^{-10}+14\,{q}^{-8}+21\,{q}^{-6}+9\,{q}^{-4}+{q}^{-22}+2\,{q}^{-20}+{q}^{-24}+{q}^{-18}+116\,{q}^{44}+13\,{q}^{42}-102\,{q}^{40}-119\,{q}^{38}-35\,{q}^{36}+83\,{q}^{34}+121\,{q}^{32}+32\,{q}^{30}-90\,{q}^{28}-157\,{q}^{26}-95\,{q}^{24}+23\,{q}^{22}+91\,{q}^{20}+64\,{q}^{18}+101\,{q}^{58}+65\,{q}^{56}-18\,{q}^{54}-72\,{q}^{52}-42\,{q}^{50}+51\,{q}^{48}+131\,{q}^{46}+29\,{q}^{72}+32\,{q}^{70}-2\,{q}^{68}-35\,{q}^{66}-36\,{q}^{64}+5\,{q}^{62}+77\,{q}^{60}-{q}^{86}+2\,{q}^{84}-10\,{q}^{82}-24\,{q}^{80}-33\,{q}^{78}-21\,{q}^{76}+10\,{q}^{74}-2\,{q}^{96}-7\,{q}^{94}-12\,{q}^{92}-13\,{q}^{90}-7\,{q}^{88}-37\,{q}^{16}-101\,{q}^{14}-89\,{q}^{12}-17\,{q}^{10}+48\,{q}^{8}+22\,{q}^{4}+55\,{q}^{6}-23\,{q}^{2} \Big) {A}^{52}+ \Big( -29-50\,{q}^{-2}-4\,{q}^{-16}-12\,{q}^{-14}-8\,{q}^{-12}+6\,{q}^{-10}+24\,{q}^{-8}+8\,{q}^{-6}-29\,{q}^{-4}-3\,{q}^{-22}-{q}^{-20}-3\,{q}^{-24}-{q}^{-26}+2\,{q}^{-18}+112\,{q}^{44}+102\,{q}^{42}-19\,{q}^{40}-144\,{q}^{38}-126\,{q}^{36}-5\,{q}^{34}+138\,{q}^{32}+173\,{q}^{30}+81\,{q}^{28}-28\,{q}^{26}-84\,{q}^{24}-18\,{q}^{22}+89\,{q}^{20}+160\,{q}^{18}+14\,{q}^{58}+56\,{q}^{56}+17\,{q}^{54}-85\,{q}^{52}-163\,{q}^{50}-135\,{q}^{48}+2\,{q}^{46}+29\,{q}^{72}+32\,{q}^{70}+21\,{q}^{68}-11\,{q}^{66}-53\,{q}^{64}-71\,{q}^{62}-48\,{q}^{60}+24\,{q}^{86}+37\,{q}^{84}+37\,{q}^{82}+16\,{q}^{80}-2\,{q}^{78}-5\,{q}^{76}+8\,{q}^{74}+7\,{q}^{98}+13\,{q}^{96}+13\,{q}^{94}+7\,{q}^{92}+{q}^{90}+7\,{q}^{88}+2\,{q}^{100}+138\,{q}^{16}+36\,{q}^{14}-46\,{q}^{12}-67\,{q}^{10}+88\,{q}^{4}+72\,{q}^{6}+44\,{q}^{2} \Big) {A}^{50}+ \Big( 64-22\,{q}^{-2}+2\,{q}^{-16}+3\,{q}^{-14}+29\,{q}^{-12}+59\,{q}^{-10}+43\,{q}^{-8}-15\,{q}^{-6}-50\,{q}^{-4}+14\,{q}^{-22}+26\,{q}^{-20}+{q}^{-24}-3\,{q}^{-26}+5\,{q}^{-28}+2\,{q}^{-36}+4\,{q}^{-34}+9\,{q}^{-32}+20\,{q}^{-18}+9\,{q}^{-30}+{q}^{-40}+{q}^{-38}+98\,{q}^{44}+221\,{q}^{42}+165\,{q}^{40}-148\,{q}^{36}-122\,{q}^{34}+{q}^{32}+92\,{q}^{30}+91\,{q}^{28}-18\,{q}^{26}-104\,{q}^{24}-151\,{q}^{22}-117\,{q}^{20}-38\,{q}^{18}-6\,{q}^{58}+79\,{q}^{56}+156\,{q}^{54}+141\,{q}^{52}+9\,{q}^{50}-93\,{q}^{48}-64\,{q}^{46}+{q}^{72}+14\,{q}^{70}+12\,{q}^{68}+24\,{q}^{66}+25\,{q}^{64}+9\,{q}^{62}-21\,{q}^{60}-30\,{q}^{86}-7\,{q}^{84}+5\,{q}^{82}-5\,{q}^{80}-32\,{q}^{78}-38\,{q}^{76}-23\,{q}^{74}-9\,{q}^{98}-5\,{q}^{96}+2\,{q}^{94}-4\,{q}^{92}-21\,{q}^{90}-36\,{q}^{88}-{q}^{104}-5\,{q}^{102}-9\,{q}^{100}+34\,{q}^{16}+58\,{q}^{14}-10\,{q}^{12}-96\,{q}^{10}-132\,{q}^{8}+50\,{q}^{4}-63\,{q}^{6}+100\,{q}^{2} \Big) {A}^{48}+ \Big( 79+40\,{q}^{-2}-59\,{q}^{-16}-28\,{q}^{-14}+4\,{q}^{-12}-24\,{q}^{-10}-94\,{q}^{-8}-128\,{q}^{-6}-73\,{q}^{-4}-10\,{q}^{-22}-25\,{q}^{-20}-20\,{q}^{-24}-36\,{q}^{-26}-36\,{q}^{-28}-7\,{q}^{-36}-{q}^{-34}-4\,{q}^{-32}-4\,{q}^{-44}-3\,{q}^{-46}-{q}^{-48}-56\,{q}^{-18}-18\,{q}^{-30}-6\,{q}^{-42}-8\,{q}^{-40}-10\,{q}^{-38}-185\,{q}^{44}+10\,{q}^{42}+137\,{q}^{40}+79\,{q}^{38}-80\,{q}^{36}-193\,{q}^{34}-111\,{q}^{32}+27\,{q}^{30}+118\,{q}^{28}+111\,{q}^{26}+35\,{q}^{24}+3\,{q}^{22}-30\,{q}^{20}-20\,{q}^{18}-90\,{q}^{58}-100\,{q}^{56}-26\,{q}^{54}+62\,{q}^{52}+35\,{q}^{50}-113\,{q}^{48}-225\,{q}^{46}-5\,{q}^{72}+8\,{q}^{70}-{q}^{68}-26\,{q}^{66}-22\,{q}^{64}-11\,{q}^{62}-38\,{q}^{60}-17\,{q}^{86}-7\,{q}^{84}+21\,{q}^{82}+29\,{q}^{80}+8\,{q}^{78}-25\,{q}^{76}-30\,{q}^{74}-5\,{q}^{98}-2\,{q}^{96}+8\,{q}^{94}+19\,{q}^{92}+15\,{q}^{90}-4\,{q}^{88}+2\,{q}^{106}+4\,{q}^{104}+4\,{q}^{102}+17\,{q}^{16}+79\,{q}^{14}+133\,{q}^{12}+74\,{q}^{10}-36\,{q}^{8}-96\,{q}^{4}-128\,{q}^{6}+18\,{q}^{2} \Big) {A}^{46}+ \Big( 20+126\,{q}^{-2}+50\,{q}^{-16}+84\,{q}^{-14}+48\,{q}^{-12}-30\,{q}^{-10}-63\,{q}^{-8}+5\,{q}^{-6}+111\,{q}^{-4}+34\,{q}^{-22}+6\,{q}^{-20}+43\,{q}^{-24}+23\,{q}^{-26}-2\,{q}^{-28}+22\,{q}^{-36}+29\,{q}^{-34}+16\,{q}^{-32}+3\,{q}^{-44}+3\,{q}^{-46}+2\,{q}^{-48}+4\,{q}^{-50}+3\,{q}^{-52}+2\,{q}^{-54}+12\,{q}^{-18}-3\,{q}^{-30}+2\,{q}^{-42}+{q}^{-40}+8\,{q}^{-38}-146\,{q}^{44}-106\,{q}^{42}+68\,{q}^{40}+175\,{q}^{38}+82\,{q}^{36}-81\,{q}^{34}-193\,{q}^{32}-120\,{q}^{30}-10\,{q}^{28}+25\,{q}^{26}-9\,{q}^{24}-82\,{q}^{22}-89\,{q}^{20}-105\,{q}^{18}+13\,{q}^{58}-33\,{q}^{56}-41\,{q}^{54}+29\,{q}^{52}+113\,{q}^{50}+99\,{q}^{48}-46\,{q}^{46}-20\,{q}^{72}+18\,{q}^{70}+42\,{q}^{68}+25\,{q}^{66}-9\,{q}^{64}-5\,{q}^{62}+17\,{q}^{60}-4\,{q}^{86}-20\,{q}^{84}-5\,{q}^{82}+29\,{q}^{80}+39\,{q}^{78}+13\,{q}^{76}-25\,{q}^{74}-{q}^{98}-6\,{q}^{96}-6\,{q}^{94}+5\,{q}^{92}+18\,{q}^{90}+15\,{q}^{88}-{q}^{108}-{q}^{106}+{q}^{104}+4\,{q}^{102}+4\,{q}^{100}-126\,{q}^{16}-111\,{q}^{14}-52\,{q}^{12}+49\,{q}^{10}+46\,{q}^{8}-140\,{q}^{4}-47\,{q}^{6}-118\,{q}^{2} \Big) {A}^{44}+ \Big( -68+42\,{q}^{-2}+52\,{q}^{-16}+25\,{q}^{-14}-48\,{q}^{-12}-71\,{q}^{-10}+112\,{q}^{-6}+139\,{q}^{-4}-14\,{q}^{-22}-14\,{q}^{-20}+12\,{q}^{-24}+18\,{q}^{-26}-2\,{q}^{-28}+19\,{q}^{-36}+8\,{q}^{-34}-15\,{q}^{-32}-4\,{q}^{-44}-{q}^{-46}+{q}^{-50}+3\,{q}^{-52}+2\,{q}^{-54}-{q}^{-58}-{q}^{-60}+20\,{q}^{-18}-21\,{q}^{-30}-8\,{q}^{-42}-4\,{q}^{-40}+11\,{q}^{-38}-18\,{q}^{44}-92\,{q}^{42}-49\,{q}^{40}+109\,{q}^{38}+199\,{q}^{36}+121\,{q}^{34}-59\,{q}^{30}+38\,{q}^{28}+138\,{q}^{26}+151\,{q}^{24}+116\,{q}^{22}+79\,{q}^{20}+110\,{q}^{18}+25\,{q}^{58}+38\,{q}^{56}-5\,{q}^{54}-16\,{q}^{52}+32\,{q}^{50}+107\,{q}^{48}+101\,{q}^{46}-45\,{q}^{72}-32\,{q}^{70}+19\,{q}^{68}+49\,{q}^{66}+23\,{q}^{64}-14\,{q}^{62}-12\,{q}^{60}+7\,{q}^{86}-14\,{q}^{84}-29\,{q}^{82}-15\,{q}^{80}+19\,{q}^{78}+25\,{q}^{76}-9\,{q}^{74}+3\,{q}^{98}-2\,{q}^{96}-8\,{q}^{94}-8\,{q}^{92}-{q}^{90}+11\,{q}^{88}-{q}^{106}-2\,{q}^{104}+3\,{q}^{100}+105\,{q}^{16}+58\,{q}^{14}+29\,{q}^{12}+71\,{q}^{10}+166\,{q}^{8}+64\,{q}^{4}+173\,{q}^{6}-60\,{q}^{2} \Big) {A}^{42}+ \Big( -105-99\,{q}^{-2}+12\,{q}^{-16}-40\,{q}^{-14}-66\,{q}^{-12}-24\,{q}^{-10}+55\,{q}^{-8}+74\,{q}^{-6}-{q}^{-4}-17\,{q}^{-22}+3\,{q}^{-20}-15\,{q}^{-24}+4\,{q}^{-26}+5\,{q}^{-28}+3\,{q}^{-36}-15\,{q}^{-34}-20\,{q}^{-32}-10\,{q}^{-44}-7\,{q}^{-46}-4\,{q}^{-48}+{q}^{-52}+2\,{q}^{-54}-2\,{q}^{-58}-2\,{q}^{-60}-{q}^{-62}+28\,{q}^{-18}-8\,{q}^{-30}-6\,{q}^{-42}+6\,{q}^{-40}+10\,{q}^{-38}+65\,{q}^{44}-25\,{q}^{42}-79\,{q}^{40}-42\,{q}^{38}+58\,{q}^{36}+106\,{q}^{34}+20\,{q}^{32}-63\,{q}^{30}-94\,{q}^{28}-21\,{q}^{26}+29\,{q}^{24}-5\,{q}^{22}-37\,{q}^{20}-51\,{q}^{18}-8\,{q}^{58}+21\,{q}^{56}+35\,{q}^{54}-4\,{q}^{52}-23\,{q}^{50}+15\,{q}^{48}+67\,{q}^{46}+5\,{q}^{72}-21\,{q}^{70}-4\,{q}^{68}+37\,{q}^{66}+51\,{q}^{64}+21\,{q}^{62}-20\,{q}^{60}+13\,{q}^{86}+11\,{q}^{84}+{q}^{82}-5\,{q}^{80}+10\,{q}^{78}+33\,{q}^{76}+34\,{q}^{74}+3\,{q}^{98}+2\,{q}^{96}-3\,{q}^{92}+5\,{q}^{88}-{q}^{104}-{q}^{102}+{q}^{100}-4\,{q}^{16}-8\,{q}^{14}-69\,{q}^{12}-93\,{q}^{10}-54\,{q}^{8}+61\,{q}^{4}+45\,{q}^{6}-18\,{q}^{2} \Big) {A}^{40}+ \Big( -23-68\,{q}^{-2}-28\,{q}^{-16}-44\,{q}^{-14}-28\,{q}^{-12}+11\,{q}^{-10}+25\,{q}^{-8}-9\,{q}^{-6}-59\,{q}^{-4}-5\,{q}^{-22}+6\,{q}^{-20}-15\,{q}^{-24}-13\,{q}^{-26}-3\,{q}^{-28}-11\,{q}^{-36}-10\,{q}^{-34}-{q}^{-32}-6\,{q}^{-46}-7\,{q}^{-48}-3\,{q}^{-50}+2\,{q}^{-52}+4\,{q}^{-54}-{q}^{-58}-2\,{q}^{-60}+2\,{q}^{-56}-{q}^{-62}-2\,{q}^{-18}+3\,{q}^{-30}+7\,{q}^{-42}+6\,{q}^{-40}-3\,{q}^{-38}-12\,{q}^{44}-17\,{q}^{42}-63\,{q}^{40}-84\,{q}^{38}-65\,{q}^{36}-9\,{q}^{34}+5\,{q}^{32}-37\,{q}^{30}-67\,{q}^{28}-70\,{q}^{26}-23\,{q}^{24}-9\,{q}^{22}-32\,{q}^{20}-41\,{q}^{18}-47\,{q}^{58}-40\,{q}^{56}-23\,{q}^{54}-22\,{q}^{52}-45\,{q}^{50}-58\,{q}^{48}-40\,{q}^{46}-13\,{q}^{72}-27\,{q}^{70}-38\,{q}^{68}-29\,{q}^{66}-11\,{q}^{64}-10\,{q}^{62}-28\,{q}^{60}-6\,{q}^{86}-8\,{q}^{84}-10\,{q}^{82}-13\,{q}^{80}-16\,{q}^{78}-15\,{q}^{76}-9\,{q}^{74}+{q}^{98}-3\,{q}^{94}-5\,{q}^{92}-4\,{q}^{90}-3\,{q}^{88}-{q}^{104}-2\,{q}^{102}-{q}^{100}-32\,{q}^{16}+2\,{q}^{14}-7\,{q}^{12}-47\,{q}^{10}-61\,{q}^{8}+20\,{q}^{4}-32\,{q}^{6}+25\,{q}^{2} \Big) {A}^{38}+ \Big( 26+9\,{q}^{-2}+{q}^{-16}+4\,{q}^{-14}+17\,{q}^{-12}+21\,{q}^{-10}+14\,{q}^{-8}+2\,{q}^{-6}-2\,{q}^{-4}+11\,{q}^{-22}+11\,{q}^{-20}+7\,{q}^{-24}+2\,{q}^{-26}+{q}^{-36}+6\,{q}^{-34}+7\,{q}^{-32}+6\,{q}^{-44}+2\,{q}^{-46}-{q}^{-48}-2\,{q}^{-50}+2\,{q}^{-54}+{q}^{-58}+2\,{q}^{-56}+3\,{q}^{-18}+5\,{q}^{-30}+4\,{q}^{-42}+{q}^{-40}-3\,{q}^{-38}+{q}^{-72}+19\,{q}^{44}+23\,{q}^{42}+28\,{q}^{40}+10\,{q}^{38}+11\,{q}^{36}+14\,{q}^{34}+24\,{q}^{32}+34\,{q}^{30}+13\,{q}^{28}+14\,{q}^{26}+12\,{q}^{24}+19\,{q}^{22}+26\,{q}^{20}+7\,{q}^{18}+12\,{q}^{58}+8\,{q}^{56}+12\,{q}^{54}+17\,{q}^{52}+19\,{q}^{50}+14\,{q}^{48}+9\,{q}^{46}+8\,{q}^{72}+9\,{q}^{70}+8\,{q}^{68}+6\,{q}^{66}+9\,{q}^{64}+13\,{q}^{62}+15\,{q}^{60}+4\,{q}^{86}+3\,{q}^{84}+{q}^{82}+5\,{q}^{80}+6\,{q}^{78}+7\,{q}^{76}+5\,{q}^{74}+{q}^{98}+2\,{q}^{96}+{q}^{94}+{q}^{90}+3\,{q}^{88}+{q}^{108}+15\,{q}^{16}+16\,{q}^{14}+24\,{q}^{12}+21\,{q}^{10}+3\,{q}^{8}+12\,{q}^{4}+3\,{q}^{6}+24\,{q}^{2} \Big) {A}^{36}$
}

\section{Conclusion}
We managed to calculate inclusive Racah matrices for $R=[3,3]$ using the highest weight method in Gelfand-Tsetlin basis. The use of this basis allows to simplify computations significantly and access colored HOMFLY polynomials with 6 and more boxes. We expect to get several new interesting results soon, in particular, the long-anticipated non-trivial color [4,2]. All details of the method will be published separately. Let us emphasis that our results helped to discover many interesting properties of quantum 6-j symbols and colored knot invariants such as new symmetries \cite{IMMMev}, a block structure of $\cal{R}$-matrix \cite{Rblocks}, relation between colored Alexander polynomial and KP hierarchy \cite{KP} and new symmetries of colored HOMFLY polynomials \cite{MSTs1, MSTs2}.

Finally, we list a few issues in which 6j symbols must necessarily play an important role, but their impact is still very poorly understood.

First, there are Khovanov-type knot polynomial invariants \cite{Kh, KhRoz} or refined invariants called superpolynomials \cite{DGR}. On the one hand, there is no R-matrix description for them, and on the other hand, its various structures show themselves perfectly \cite{AnMor, AnMorPop}.  In addition, in \cite{ArtSh1, ArtSh2}, analogs of the squares of 6j-symbols for superpolynomials of the DAHA algebra for knots of genus 2 were obtained. This indicates that refined analogs of quantum 6j symbols are also present in the description of colored superpolynomials. 

Second, it is a matrix-model approach or a topological recursion method. These two methods are closely related to each other. It is well known that matrix models can describe various combinatorial quantities often associated with graphs (see, for example, recent paper \cite{AmOrVa}). Therefore, it is natural to expect their appearance in knot theory. Topological recursion is, in a sense, an algebraic-geometric interpretation of the loop equations of the matrix model, which, however, may turn out to be a more general construction \cite{toprec}. In any case, in the context of colored HOMFLY polynomials, both approaches are developed only for torus knots \cite{BEM,DBPSS}, where the role of 6j-symbols is negligible. Nevertheless, it is believed that for HOMFLY polynomials of non-torus knots, if there is a matrix model, then it necessarily takes into account the influence of 6j-symbols. It is also possible that instead of matrix models one should consider more complex tensor models, which, incidentally, have all the properties necessary for this task: character expansion \cite{MirMortenz}, diagram technique \cite{AmVas}, etc.

Third, 6j-symbols appear in quantum computing. Moreover, both classical 6j-symbols can appear when calculating the wave function of a quantum system \cite{quantcomp}, and quantum when calculating R-matrices \cite{KolMor} in the topological approach \cite{MMMM}.

\section*{Acknowledgements}
We are indebted to Alexei Morozov for a lot of discussions on polynomial invariants and Racah matrices. Shamil Shakirov is also grateful to Nicolai Reshetikhin for enlightening explanations on quantum groups and 6-j symbols. 

This work was funded by the Russian Science Foundation (Grant No. 20-71-10073).

\pagebreak

\section{Appendix A: Example of computation}

To give a more detailed illustration of our method, let us present here the computation of the Racah matrix for $Q = [9, 5, 4]$. This is a simple case, since the size of Racah matrix is 2.

The essence of computation is comparison of two different bases in one and the same vector space of intertwiners $[3,3] \otimes [3,3] \otimes [3,3] \rightarrow [9,5,4]$. The Racah matrix arises as the matrix of reexpansion coefficients between one basis and another. It is clear from general representation theory perspective that full information about these bases is contained in their highest weight components -- the preimages under the intertwining map of the highest weight in representation $[9,5,4]$. These preimages represent two sets of vectors in the tensor cube $[3,3] \otimes [3,3] \otimes [3,3]$, labeled by intermediate choices of internal representation in the double tensor product $[3,3] \otimes [3,3]$. For computation of these vectors we employ a standard representation theory technique, using the action of $U_q(sl_N)$ for suitable $N$ (here we choose $N = 3$ as the target Young diagram has 3 rows) on tensor products of irreducible finite-dimensional representations. 

The crucial new feature that we use in this paper is the use of a distinguished basis for these finite-dimensional representations: the basis of Gelfand-Tseitlin tables, which in this case for $N = 3$, represent 6 integers arranged in a form

$$
\big| M \big> =
\left(\begin{array}{ccccccc}
\lambda_1 \ \ \ \ \ \lambda_2 \ \ \ \ \ \lambda_3 \\
\\
m_{1,1} \ \ \ m_{2,1} \\
\\
m_{2,2}
\end{array}\right)
$$
\smallskip\\
that satisfy the betweenness conditions

$$
\lambda_1 \geq m_{1,1} \geq \lambda_2, \ \ \ \lambda_2 \geq m_{2,1} \geq \lambda_3, \ \ \ m_{1,1} \geq m_{2,2} \geq m_{2,1}
$$
\smallskip\\
and label basis vectors in a highest weight representation of $U_q(sl_3)$ corresponding to a Young diagram $\lambda = (\lambda_1 \geq \lambda_2 \geq \lambda_3)$. This basis is distinguished by several properties, importantly it enjoys nice and simple formulas for the action of the algebra. This is what leads to drastic simplification of the whole computation as compared to the previous realizations of the method in \cite{MMMS21,MMMS31,MMMS22} and allows to consider more complicated colors of knots and links.

Direct computation using the action of $U_q(sl_3)$ on $[3,3] \otimes [3,3] \otimes [3,3]$ leads to the following expressions for the sets of vectors in question. For the left basis we have

$$
v^{\rm{left}}_{[6,5,1]} = \dfrac{q^2}{1+q^4} e_1 \otimes e_0 \otimes e_3 - \dfrac{q^2(1+q^2)}{1+q^4} e_2 \otimes e_0 \otimes e_2 + e_3 \otimes e_0 \otimes e_1 + \ldots
$$
$$
v^{\rm{left}}_{[6,4,2]} = - \dfrac{1+q^2+q^4}{q^2(1+q^2)} e_1 \otimes e_0 \otimes e_3 + e_2 \otimes e_0 \otimes e_2 + 0 \ e_3 \otimes e_0 \otimes e_1 + \ldots
$$
\smallskip\\
and for the right basis we have

$$
v^{\rm{right}}_{[6,5,1]} = e_1 \otimes e_0 \otimes e_3 - \dfrac{q^2(1+q^2)}{1+q^4} e_2 \otimes e_0 \otimes e_2 + \dfrac{q^2}{1+q^4} e_3 \otimes e_0 \otimes e_1 + \ldots
$$
$$
v^{\rm{right}}_{[6,4,2]} = 0 \ e_1 \otimes e_0 \otimes e_3 + e_2 \otimes e_0 \otimes e_2 - \dfrac{1+q^2+q^4}{q^2(1+q^2)} e_3 \otimes e_0 \otimes e_1 + \ldots
$$
\smallskip\\
where

$$
e_k = \left(\begin{array}{ccccccc}
3 \ \ \ 3 \ \ \ 0 \\
\\
3 \ \ \ 0 \\
\\
k
\end{array}\right), \ \ \ k = 0,1,2,3
$$
\smallskip\\
are certain vectors in the Gelfand-Tseitlin basis for representation $[3,3]$. Note that the vectors are labeled by intermediate representations in the double tensor product $[3,3] \otimes [3,3]$, which can be only $[6,5,1]$ or $[6,4,2]$ for the case of target $Q = [9, 5, 4]$ that we currently consider. Note also that we do not write out the full list of components of the vectors, since this is not necessary: for the purpose of determining the $2 \times 2$ matrix that we look for, it is sufficient to only consider the components where the second tensor factor is fixed to be the highest weight vector $e_0$. This is another advantage of the Gelfand-Tseitlin realization of the method of \cite{MMMS21,MMMS31,MMMS22}: we save time and resource by calculating only a small part of the vectors in question.

Expanding the left set of vectors in the basis of the right set of vectors, we find a $2 \times 2$ reexpansion matrix

$$
\left(\begin{array}{c}
v^{\rm{left}}_{[6,5,1]} \\
\\
v^{\rm{left}}_{[6,4,2]}
\end{array}\right) =
\underbrace{ \left(\begin{array}{cc}
	\dfrac{q^2}{1+q^4} & \dfrac{-q^2(1+q^6)}{(1+q^4)^2} \\
	\\
	\dfrac{-(1+q^2+q^4)}{q^2(1+q^2)} & \dfrac{-q^2}{1+q^4}
	\end{array}\right)}_{U} \cdot
\left(\begin{array}{c}
v^{\rm{right}}_{[6,5,1]} \\
\\
v^{\rm{right}}_{[6,4,2]}
\end{array}\right)
$$
\smallskip\\
The matrix that we obtained is not orthogonal, which is not surprising since we used arbitrary normalization of both left and right sets of vectors. In order to obtain the correct Racah matrix this $2 \times 2$ matrix needs to be orthogonalized. For most efficient implementation of this step we do not need to renormalize the left and right vectors, but rather work with the matrix itself. Denoting the $2 \times 2$ matrix we obtained as $U$, we look for diagonal $2 \times 2$ matrices $V_1$ and $V_2$ such that $V_1 U V_2$ would be orthogonal. This is easy to do and solution is unique modulo some trivial remainder freedom. The resulting Racah matrix is

$$
{\cal U} = V_1 U V_2 = \left(\begin{array}{cc}
\dfrac{q^2}{1+q^4} & \dfrac{-\sqrt{1+q^4+q^8}}{1+q^4} \\
\\
\dfrac{-\sqrt{1+q^4+q^8}}{1+q^4} & \dfrac{-q^2}{1+q^4}
\end{array}\right)
$$
\smallskip\\
and this is the right answer, as easily checked by the eigenvalue conjecture. This concludes our illustration of the very promising Gelfand-Tseitlin version of the highest weight method, which hopefully showcases the main advantages of working in Gelfand-Tseitlin basis.

\section*{Appendix B: Matrix elements of the Racah matrix $[3,3]^{\otimes 3} \rightarrow [7,6,3,2]$}
\emph{}
Here we present for illustration purposes the irreducible factors of the $12 \times 12$ Racah matrix $[3,3]^{\otimes 3} \rightarrow [7,6,3,2]$.

{\fontsize{4pt}{0pt}

\begin{center}
$
u_1^2 = u_1^3 = - {q}^{10} \ \big( {q}^{40}+2\,{q}^{38}+3\,{q}^{36}+6\,{q}^{34}
+8\,{q}^{32}+11\,{q}^{30}+14\,{q}^{28}+17\,{q}^{26}+19\,{q}^{24}+20\,{
q}^{22}+21\,{q}^{20}+20\,{q}^{18}+19\,{q}^{16}+17\,{q}^{14}+14\,{q}^{
12}+11\,{q}^{10}+8\,{q}^{8}+6\,{q}^{6}+3\,{q}^{4}+2\,{q}^{2}+1 \big)^{-1/2}
$
\end{center} \vspace{-4ex} \begin{center}
$
u_1^4 = - u_1^5 = {q}^{11} \ \big( {q}^{44}+3\,{q}^{42}+7\,{q}^{40}+13\,{q}^{38}
+21\,{q}^{36}+32\,{q}^{34}+43\,{q}^{32}+54\,{q}^{30}+66\,{q}^{28}+73\,
{q}^{26}+78\,{q}^{24}+81\,{q}^{22}+78\,{q}^{20}+73\,{q}^{18}+66\,{q}^{
16}+54\,{q}^{14}+43\,{q}^{12}+32\,{q}^{10}+21\,{q}^{8}+13\,{q}^{6}+7\,
{q}^{4}+3\,{q}^{2}+1 \big)^{-1/2}
$
\end{center} \vspace{-4ex} \begin{center}
$
u_1^8 = u_1^9 = - u_8^1 = - u_9^1 = {q}^{9} \ \big( {q}^{36}+4\,{q}^{34}+9\,{q}^{32}+15\,{q}^{30}+
23\,{q}^{28}+32\,{q}^{26}+41\,{q}^{24}+48\,{q}^{22}+52\,{q}^{20}+53\,{
q}^{18}+52\,{q}^{16}+48\,{q}^{14}+41\,{q}^{12}+32\,{q}^{10}+23\,{q}^{8
}+15\,{q}^{6}+9\,{q}^{4}+4\,{q}^{2}+1 \big)^{-1/2}
$
\end{center} \vspace{-4ex} \begin{center}
$
u_2^1 = - u_3^1 = -{q}^{7} \ \big( {q}^{28}+4\,{q}^{24}+{q}^{22}+7\,{q}^{20}+{q}
^{18}+9\,{q}^{16}+{q}^{14}+9\,{q}^{12}+{q}^{10}+7\,{q}^{8}+{q}^{6}+4\,
{q}^{4}+1 \big)^{-1/2}
$
\end{center} \vspace{-4ex} \begin{center}
$
u_4^1 = - {q}^{6} \left( {q}^{12}+3\,{q}^{10}+3\,{q}^{8}+3\,{q}^{6}+3\,{q}^{4}
+3\,{q}^{2}+1 \right) \ \big( {q}^{48}+6\,{q}^{46}+18\,{q}^{44
}+39\,{q}^{42}+72\,{q}^{40}+119\,{q}^{38}+178\,{q}^{36}+241\,{q}^{34}+
308\,{q}^{32}+366\,{q}^{30}+413\,{q}^{28}+440\,{q}^{26}+452\,{q}^{24}+
440\,{q}^{22}+413\,{q}^{20}+366\,{q}^{18}+308\,{q}^{16}+241\,{q}^{14}+
178\,{q}^{12}+119\,{q}^{10}+72\,{q}^{8}+39\,{q}^{6}+18\,{q}^{4}+6\,{q}
^{2}+1 \big)^{-1/2}
$
\end{center} \vspace{-4ex} \begin{center}
$
u_5^1 = {q}^{12} \left( {q}^{12}+3\,{q}^{10}+3\,{q}^{8}+3\,{q}^{6}+3\,{q}^{4}
+3\,{q}^{2}+1 \right) \ \big( {q}^{48}+6\,{q}^{46}+18\,{q}^{44}+39\,{q}^{42
}+72\,{q}^{40}+119\,{q}^{38}+178\,{q}^{36}+241\,{q}^{34}+308\,{q}^{32}
+366\,{q}^{30}+413\,{q}^{28}+440\,{q}^{26}+452\,{q}^{24}+440\,{q}^{22}
+413\,{q}^{20}+366\,{q}^{18}+308\,{q}^{16}+241\,{q}^{14}+178\,{q}^{12}
+119\,{q}^{10}+72\,{q}^{8}+39\,{q}^{6}+18\,{q}^{4}+6\,{q}^{2}+1 \big)^{-1/2}
$
\end{center} \vspace{-4ex} \begin{center}
$
u_{2}^{2}=251 + 5{q}^{22} + 13{q}^{20} + 48{q}^{-16} + 171{q}^{-8} + 13
{q}^{-20} + {q}^{24} + 171{q}^{8} + 244{q}^{-2} + 244{q}^{2} + 5{q}^{-
22} + 75{q}^{14} + 105{q}^{12} + 28{q}^{18} + 138{q}^{10} + 48{q}^{16}
 + {q}^{-24} + 28{q}^{-18} + 75{q}^{-14} + 105{q}^{-12} + 138{q}^{-10} +
202{q}^{6} + 202{q}^{-6} + 228{q}^{4} + 228{q}^{-4}
$ \end{center} \vspace{-4ex} \begin{center} $
u_{2}^{3}={q}^{10} + {q}^{8} + {q}^{6} + 2{q}^{4} + 2{q}^{2} + 2{q}^{-2} + 2
{q}^{-4} + {q}^{-6} + {q}^{-8} + {q}^{-10}
$ \end{center} \vspace{-4ex} \begin{center} $
u_{2}^{4}=463 + {q}^{26} + 15{q}^{22} + 35{q}^{20} + 106{q}^{-16} + 323{
q}^{-8} + 35{q}^{-20} + {q}^{-26} + 5{q}^{24} + 323{q}^{8} + 453{q}^{-2}
 + 453{q}^{2} + 15{q}^{-22} + 155{q}^{14} + 209{q}^{12} + 65{q}^{18} +
266{q}^{10} + 106{q}^{16} + 5{q}^{-24} + 65{q}^{-18} + 155{q}^{-14} +
209{q}^{-12} + 266{q}^{-10} + 377{q}^{6} + 377{q}^{-6} + 422{q}^{4} +
422{q}^{-4}
$ \end{center} \vspace{-4ex} \begin{center} $
u_{2}^{5}=-({q}^{12} + 2{q}^{10} + 3{q}^{8} + 3{q}^{6} + 4{q}^{4} + 3{q}
^{2} + 2 + 3{q}^{-2} + 4{q}^{-4} + 3{q}^{-6} + 3{q}^{-8} + 2{q}^{-10} + {q
}^{-12})
$ \end{center} \vspace{-4ex} \begin{center} $
u_{2}^{6}={q}^{10} + 2{q}^{8} + {q}^{6} + 2{q}^{4} + {q}^{2} + 2 + {q}^{-2} + 2
{q}^{-4} + {q}^{-6} + 2{q}^{-8} + {q}^{-10}
$ \end{center} \vspace{-4ex} \begin{center} $
u_{2}^{7}=-({q}^{18} + 2{q}^{16} + 3{q}^{14} + 5{q}^{12} + 6{q}^{10} + 7
{q}^{8} + 9{q}^{6} + 9{q}^{4} + 11{q}^{2} + 11 + 11{q}^{-2} + 9{q}^{-4} +
9{q}^{-6} + 7{q}^{-8} + 6{q}^{-10} + 5{q}^{-12} + 3{q}^{-14} + 2{q}^
{-16} + {q}^{-18})
$ \end{center} \vspace{-4ex} \begin{center} $
u_{2}^{8}={q}^{18} + {q}^{16} + 2{q}^{14} + {q}^{12} + 3{q}^{10} + 2{q}^{8
} + 3{q}^{6} + 2{q}^{4} + 4{q}^{2} + 3 + 4{q}^{-2} + 2{q}^{-4} + 3{q}^{-
6} + 2{q}^{-8} + 3{q}^{-10} + {q}^{-12} + 2{q}^{-14} + {q}^{-16} + {q}^{-18}
$ \end{center} \vspace{-4ex} \begin{center} $
u_{2}^{9}=31 + 3{q}^{22} + 5{q}^{20} + 11{q}^{-16} + 25{q}^{-8} + 5{q}
^{-20} + {q}^{24} + 25{q}^{8} + 30{q}^{-2} + 30{q}^{2} + 3{q}^{-22} + 14
{q}^{14} + 18{q}^{12} + 8{q}^{18} + 21{q}^{10} + 11{q}^{16} + {q}^{-24} +
8{q}^{-18} + 14{q}^{-14} + 18{q}^{-12} + 21{q}^{-10} + 28{q}^{6} + 28
{q}^{-6} + 30{q}^{4} + 30{q}^{-4}
$ \end{center} \vspace{-4ex} \begin{center} $
u_{2}^{10}=-{q}^{12} + {q}^{10} + {q}^{2}-1 + {q}^{-2} + {q}^{-10}-{q}^{-12}
$ \end{center} \vspace{-4ex} \begin{center} $
u_{2}^{11}=-{q}^{14} + 2{q}^{12}-3{q}^{10} + 5{q}^{8}-5{q}^{6} + 8
{q}^{4}-7{q}^{2} + 9-7{q}^{-2} + 8{q}^{-4}-5{q}^{-6} + 5{q}^{-8}-3
{q}^{-10} + 2{q}^{-12}-{q}^{-14}
$ \end{center} \vspace{-4ex} \begin{center} $
u_{2}^{12}={q}^{14} + 4{q}^{10} + {q}^{8} + 7{q}^{6} + {q}^{4} + 9{q}^{2} +
1 + 9{q}^{-2} + {q}^{-4} + 7{q}^{-6} + {q}^{-8} + 4{q}^{-10} + {q}^{-14}
$ \end{center} \vspace{-4ex} \begin{center} $
u_{3}^{2} = -{q}^{8}-{q}^{6}-{q}^{4} + 1-{q}^{-4}-{q}^{-6}-{q}^{-8}
$ \end{center} \vspace{-4ex} \begin{center} $
u_{3}^{3} = 2 {q}^{16} + 4 {q}^{14} + 8 {q}^{12} + 12 {q}^{10} + 16 {q}^{8}
 + 19 {q}^{6} + 22 {q}^{4} + 25 {q}^{2} + 25 + 25 {q}^{-2} + 22 {q}^{-4} + 19
{q}^{-6} + 16 {q}^{-8} + 12 {q}^{-10} + 8 {q}^{-12} + 4 {q}^{-14} + 2 {q}^{
-16}
$ \end{center} \vspace{-4ex} \begin{center} $
u_{3}^{4}=-({q}^{14} + 3{q}^{12} + 6{q}^{10} + 10{q}^{8} + 12{q}^{6} + 14
{q}^{4} + 15{q}^{2} + 15 + 15{q}^{-2} + 14{q}^{-4} + 12{q}^{-6} + 10{q
}^{-8} + 6{q}^{-10} + 3{q}^{-12} + {q}^{-14})
$ \end{center} \vspace{-4ex} \begin{center} $
u_{3}^{5}=2{q}^{18} + 7{q}^{16} + 16{q}^{14} + 28{q}^{12} + 42{q}^{
10} + 55{q}^{8} + 66{q}^{6} + 75{q}^{4} + 81{q}^{2} + 83 + 81{q}^{-2} + 75
{q}^{-4} + 66{q}^{-6} + 55{q}^{-8} + 42{q}^{-10} + 28{q}^{-12} + 16{
q}^{-14} + 7{q}^{-16} + 2{q}^{-18}
$ \end{center} \vspace{-4ex} \begin{center} $
u_{3}^{7}=-2{q}^{4}-4{q}^{2}-5-4{q}^{-2}-2{q}^{-4}
$ \end{center} \vspace{-4ex} \begin{center} $
u_{3}^{9}=3{q}^{16} + 11{q}^{14} + 25{q}^{12} + 42{q}^{10} + 59{q}^{
8} + 74{q}^{6} + 88{q}^{4} + 99{q}^{2} + 103 + 99{q}^{-2} + 88{q}^{-4} +
74{q}^{-6} + 59{q}^{-8} + 42{q}^{-10} + 25{q}^{-12} + 11{q}^{-14} + 3
{q}^{-16}
$ \end{center} \vspace{-4ex} \begin{center} $
u_{4}^{2} = {q}^{-18} \big( {q}^{48} + 2{q}^{46} + 5{q}^{44} + 6{q}^{42} + 14{q}
^{40} + 10{q}^{38} + 24{q}^{36} + 15{q}^{34} + 34{q}^{32} + 20{q}^{30}
 + 43{q}^{28} + 23{q}^{26} + 47{q}^{24} + 23{q}^{22} + 43{q}^{20} + 20
{q}^{18} + 34{q}^{16} + 15{q}^{14} + 24{q}^{12} + 10{q}^{10} + 14{q}^{
8} + 6{q}^{6} + 5{q}^{4} + 2{q}^{2} + 1 \big) \ \big( {q}^{12} + 3{q}
^{10} + 3{q}^{8} + 3{q}^{6} + 3{q}^{4} + 3{q}^{2} + 1 \big)^{-1}
$ \end{center} \vspace{-4ex} \begin{center} $
u_{4}^{3}= \big( {q}^{28} + 4{q}^{26} + 6{q}^{24} + 8{q}^{22} + 11{q}
^{20} + 11{q}^{18} + 11{q}^{16} + 13{q}^{14} + 11{q}^{12} + 11{q}^{10}
 + 11{q}^{8} + 8{q}^{6} + 6{q}^{4} + 4{q}^{2} + 1 \big) \times \newline
\times {q}^{-8} \big( {q}^{12} + 3{q}^{10} + 3{q}^{8} + 3{q}^{6} + 3{q}^{4} + 3{q}^{2} + 1 \big)^{-1}
$ \end{center} \vspace{-4ex} \begin{center} $
u_{4}^{4} = {q}^{-26} \big( {q}^{64} + 4{q}^{62} + 8{q}^{60} + 16{q}^{58} + 18{q
}^{56} + 19{q}^{54} + {q}^{52} - 28{q}^{50} - 84{q}^{48} - 153{q}^{46} -
248{q}^{44} - 344{q}^{42} - 455{q}^{40} - 544q^38 - 628{q}^{36} -
673{q}^{34} - 695{q}^{32} - 673{q}^{30} - 628{q}^{28} - 544{q}^{26} -
455{q}^{24} - 344{q}^{22} - 248{q}^{20} - 153{q}^{18} - -84{q}^{16} -
28{q}^{14} + {q}^{12} + 19{q}^{10} + 18{q}^{8} + 16{q}^{6} + 8{q}^{4} +
4{q}^{2} + 1\big)
\ \big( {q}^{12} + 3{q}^{10} + 3{q}^{8} + 3{q}^{6} + 3{q}^{4} + 3{q}^{2} + 1 \big)^{-1}
$ \end{center} \vspace{-4ex} \begin{center} $
u_{4}^{5} = {q}^{-16} \big( {q}^{44} + 6{q}^{42} + 17{q}^{40} + 33{q}^{38} + 51{
q}^{36} + 70{q}^{34} + 89{q}^{32} + 109{q}^{30} + 129{q}^{28} + 147{q}
^{26} + 160{q}^{24} + 165{q}^{22} + 160{q}^{20} + 147{q}^{18} + 129{q}
^{16} + 109{q}^{14} + 89{q}^{12} + 70{q}^{10} + 51{q}^{8} + 33{q}^{6} +
17{q}^{4} + 6{q}^{2} + 1 \big)
\big( {q}^{12} + 3{q}^{10} + 3{q}^{8} + 3{q}^{6} + 3{q}^{4} + 3{q}^{2} + 1 \big)^{-1}
$ \end{center} \vspace{-4ex} \begin{center} $
u_{4}^{6} = {q}^{-6} \big( {q}^{24} + {q}^{22} + 2{q}^{20} + 5{q}^{18} + 8{q}^{16
} + 12{q}^{14} + 15{q}^{12} + 12{q}^{10} + 8{q}^{8} + 5{q}^{6} + 2{q}^
{4} + {q}^{2} + 1 \big)
\ \big( {q}^{12} + 3{q}^{10} + 3{q}^{8} + 3{q}^{6} + 3{q}^{4} + 3{q}^{2} + 1 \big)^{-1}
$ \end{center} \vspace{-4ex} \begin{center} $
u_{4}^{7}=- {q}^{-14} \big( {q}^{40} + 2{q}^{38} + 3{q}^{36} + 4{q}^{34} + 3{q}
^{32} + 5{q}^{30} + 2{q}^{28} + 6{q}^{26} + 2{q}^{24} + 7{q}^{22} + 3{
q}^{20} + 7{q}^{18} + 2{q}^{16} + 6{q}^{14} + 2{q}^{12} + 5{q}^{10} + 3
{q}^{8} + 4{q}^{6} + 3{q}^{4} + 2{q}^{2} + 1 \big)
\ \big( {q}^{12} + 3{q}^{10} + 3{q}^{8} + 3{q}^{6} + 3{q}^{4} + 3{q}^{2} + 1 \big)^{-1}
$ \end{center} \vspace{-4ex} \begin{center} $
u_{4}^{8}=- {q}^{-18} \big( {q}^{48} + 3{q}^{46} + 7{q}^{44} + 16{q}^{42} + 28{
q}^{40} + 44{q}^{38} + 62{q}^{36} + 84{q}^{34} + 105{q}^{32} + 124{q}^
{30} + 140{q}^{28} + 151{q}^{26} + 153{q}^{24} + 151{q}^{22} + 140{q}^
{20} + 124{q}^{18} + 105{q}^{16} + 84{q}^{14} + 62{q}^{12} + 44{q}^{10
} + 28{q}^{8} + 16{q}^{6} + 7{q}^{4} + 3{q}^{2} + 1 \big)
\ \big( {q}^{12} + 3{q}^{10} + 3{q}^{8} + 3{q}^{6} + 3{q}^{4} + 3{q}^{2} + 1 \big)^{-1}
$ \end{center} \vspace{-4ex} \begin{center} $
u_{4}^{9} = {q}^{-18} \big( {q}^{48} + 2{q}^{46} + 2{q}^{44} + 3{q}^{42} + {q}^{40
} + 3{q}^{38} + {q}^{36} + {q}^{34} + 2{q}^{32}-{q}^{30}-{q}^{26}-3{q}^{
24}-{q}^{22}-{q}^{18} + 2{q}^{16} + {q}^{14} + {q}^{12} + 3{q}^{10} + {q}^{8
} + 3{q}^{6} + 2{q}^{4} + 2{q}^{2} + 1 \big)
\ \big( {q}^{12} + 3{q}^{10} + 3{q}^{8} + 3{q}^{6} + 3{q}^{4} + 3{q}^{2} + 1 \big)^{-1}
$ \end{center} \vspace{-4ex} \begin{center} $
u_{4}^{10}= {q}^{-16} \big( {q}^{44} + 2{q}^{42} + {q}^{40}-2{q}^{36}-5{q}^{
34}-12{q}^{32}-18{q}^{30}-22{q}^{28}-24{q}^{26}-26{q}^{24}-
25{q}^{22}-26{q}^{20}-24{q}^{18}-22{q}^{16}-18{q}^{14}-12{
q}^{12}-5{q}^{10}-2{q}^{8} + {q}^{4} + 2{q}^{2} + 1 \big)
\ \big( {q}^{12} + 3{q}^{10} + 3{q}^{8} + 3{q}^{6} + 3{q}^{4} + 3{q}^{2} + 1 \big)^{-1}
$ \end{center} \vspace{-4ex} \begin{center} $
u_{4}^{11}= {q}^{-14} \big( {q}^{40} + 3{q}^{38} + 6{q}^{36} + 10{q}^{34} + 13{
q}^{32} + 20{q}^{30} + 22{q}^{28} + 29{q}^{26} + 29{q}^{24} + 35{q}^{
22} + 32{q}^{20} + 35{q}^{18} + 29{q}^{16} + 29{q}^{14} + 22{q}^{12} +
20{q}^{10} + 13{q}^{8} + 10{q}^{6} + 6{q}^{4} + 3{q}^{2} + 1 \big)
\ \big( {q}^{12} + 3{q}^{10} + 3{q}^{8} + 3{q}^{6} + 3{q}^{4} + 3{q}^{2} + 1 \big)^{-1}
$ \end{center} \vspace{-4ex} \begin{center} $
u_{4}^{12} = - {q}^{-12} \big( {q}^{36} + 3{q}^{34} + 5{q}^{32} + 9{q}^{30} + 13{
q}^{28} + 19{q}^{26} + 21{q}^{24} + 26{q}^{22} + 26{q}^{20} + 28{q}^{
18} + 26{q}^{16} + 26{q}^{14} + 21{q}^{12} + 19{q}^{10} + 13{q}^{8} + 9
{q}^{6} + 5{q}^{4} + 3{q}^{2} + 1 \big) \ \big( {q}^{12} + 3{q}^{10
} + 3{q}^{8} + 3{q}^{6} + 3{q}^{4} + 3{q}^{2} + 1 \big)^{-1}
$ \end{center} \vspace{-4ex} \begin{center} $
u_{5}^{2}={q}^{16} + 5{q}^{14} + 13{q}^{12} + 24{q}^{10} + 36{q}^{
8} + 47{q}^{6} + 55{q}^{4} + 61{q}^{2} + 63 + 61{q}^{-2} + 55{q}^{-4} + 47
{q}^{-6} + 36{q}^{-8} + 24{q}^{-10} + 13{q}^{-12} + 5{q}^{-14} + {q}^{
-16}
$ \end{center} \vspace{-4ex} \begin{center} $
u_{5}^{3}={q}^{22} + {q}^{20} + {q}^{18} + {q}^{16}-2{q}^{14}-4{q}^{
12}-8{q}^{10}-8{q}^{8}-15{q}^{6}-12{q}^{4}-18{q}^{2}-15-18
{q}^{-2}-12{q}^{-4}-15{q}^{-6}-8{q}^{-8}-8{q}^{-10}-4{q}^{-
12}-2{q}^{-14} + {q}^{-16} + {q}^{-18} + {q}^{-20} + {q}^{-22}
$ \end{center} \vspace{-4ex} \begin{center} $
u_{5}^{4}=333 + 6{q}^{22} + 18{q}^{20} + 67{q}^{-16} + 225{q}^{-8}
 + 18{q}^{-20} + {q}^{24} + 225{q}^{8} + 325{q}^{-2} + 325{q}^{2} + 6{q}
^{-22} + 102{q}^{14} + 140{q}^{12} + 38{q}^{18} + 182{q}^{10} + 67{q}^
{16} + {q}^{-24} + 38{q}^{-18} + 102{q}^{-14} + 140{q}^{-12} + 182{q}^{-
10} + 265{q}^{6} + 265{q}^{-6} + 301{q}^{4} + 301{q}^{-4}
$ \end{center} \vspace{-4ex} \begin{center} $
u_{5}^{5}=-61 + {q}^{28} + 2{q}^{26}-3{q}^{20}-16{q}^{-16}-43{
q}^{-8}-3{q}^{-20} + 2{q}^{-26} + {q}^{-28} + 2{q}^{24}-43{q}^{8}-60
{q}^{-2}-60{q}^{2}-21{q}^{14}-27{q}^{12}-9{q}^{18}-35{q}^{
10}-16{q}^{16} + 2{q}^{-24}-9{q}^{-18}-21{q}^{-14}-27{q}^{-12}
-35{q}^{-10}-49{q}^{6}-49{q}^{-6}-57{q}^{4}-57{q}^{-4}
$ \end{center} \vspace{-4ex} \begin{center} $
u_{5}^{6}={q}^{14} + 2{q}^{12} + 2{q}^{10} + {q}^{8} + 3{q}^{6} + 2{
q}^{4} + 4{q}^{2} + 3 + 4{q}^{-2} + 2{q}^{-4} + 3{q}^{-6} + {q}^{-8} + 2{q
}^{-10} + 2{q}^{-12} + {q}^{-14}
$ \end{center} \vspace{-4ex} \begin{center} $
u_{5}^{7}=2{q}^{12} + 7{q}^{10} + 13{q}^{8} + 18{q}^{6} + 22{q}^
{4} + 23{q}^{2} + 23 + 23{q}^{-2} + 22{q}^{-4} + 18{q}^{-6} + 13{q}^{-8}
 + 7{q}^{-10} + 2{q}^{-12}
$ \end{center} \vspace{-4ex} \begin{center} $
u_{5}^{8}=-{q}^{6}-2{q}^{4} + {q}^{2}-2 + {q}^{-2}-2{q}^{-4}-{q}^{
-6}
$ \end{center} \vspace{-4ex} \begin{center} $
u_{5}^{9}=-(79 + 10{q}^{22} + 25{q}^{20} + 62{q}^{-16} + 87{q}^{-8} +
25{q}^{-20} + 2{q}^{24} + 87{q}^{8} + 81{q}^{-2} + 81{q}^{2} + 10{q}
^{-22} + 74{q}^{14} + 80{q}^{12} + 44{q}^{18} + 84{q}^{10} + 62{q}^{16
} + 2{q}^{-24} + 44{q}^{-18} + 74{q}^{-14} + 80{q}^{-12} + 84{q}^{-10}
 + 87{q}^{6} + 87{q}^{-6} + 85{q}^{4} + 85{q}^{-4})
$ \end{center} \vspace{-4ex} \begin{center} $
u_{5}^{10}={q}^{8} + 3{q}^{6} + {q}^{4} + 2{q}^{2} + 2 + 2{q}^{-2} + {q}
^{-4} + 3{q}^{-6} + {q}^{-8}
$ \end{center} \vspace{-4ex} \begin{center} $
u_{5}^{11}=-2{q}^{4}-2{q}^{2}-3-2{q}^{-2}-2{q}^{-4}
$ \end{center} \vspace{-4ex} \begin{center} $
u_{6}^{2}={q}^{8}+2\,{q}^{4}+{q}^{2}+3+{q}^{-2}+2\,{q}^{-4}+{q}^{-
8}
$ \end{center} \vspace{-4ex} \begin{center} $
u_{6}^{3}=-{q}^{10}-2\,{q}^{8}-2\,{q}^{6}-2\,{q}^{4}-2\,{q}^{2}-1-
2\,{q}^{-2}-2\,{q}^{-4}-2\,{q}^{-6}-2\,{q}^{-8}-{q}^{-10}
$ \end{center} \vspace{-4ex} \begin{center} $
u_{6}^{4}={q}^{16}+{q}^{14}+2\,{q}^{12}+2\,{q}^{10}+2\,{q}^{8}+2\,
{q}^{6}+{q}^{4}-{q}^{2}-1-{q}^{-2}+{q}^{-4}+2\,{q}^{-6}+2\,{q}^{-8}+2
\,{q}^{-10}+2\,{q}^{-12}+{q}^{-14}+{q}^{-16}
$ \end{center} \vspace{-4ex} \begin{center} $
u_{6}^{5}=-{q}^{16}-3\,{q}^{14}-5\,{q}^{12}-5\,{q}^{10}-5\,{q}^{8}
-6\,{q}^{6}-8\,{q}^{4}-10\,{q}^{2}-11-10\,{q}^{-2}-8\,{q}^{-4}-6\,{q}^
{-6}-5\,{q}^{-8}-5\,{q}^{-10}-5\,{q}^{-12}-3\,{q}^{-14}-{q}^{-16}
$ \end{center} \vspace{-4ex} \begin{center} $
u_{6}^{6}={q}^{12}+{q}^{10}-{q}^{6}+3\,{q}^{2}+5+3\,{q}^{-2}-{q}^{
-6}+{q}^{-10}+{q}^{-12}
$ \end{center} \vspace{-4ex} \begin{center} $
u_{6}^{8}=-{q}^{8}-{q}^{6}-2\,{q}^{4}-3\,{q}^{2}-3-3\,{q}^{-2}-2\,
{q}^{-4}-{q}^{-6}-{q}^{-8}
$ \end{center} \vspace{-4ex} \begin{center} $
u_{6}^{9}={q}^{4}-{q}^{2}+1-{q}^{-2}+{q}^{-4}
$ \end{center} \vspace{-4ex} \begin{center} $
u_{6}^{10}={q}^{6}-{q}^{2}-1-{q}^{-2}+{q}^{-6}
$ \end{center} \vspace{-4ex} \begin{center} $
u_{7}^{2}=-{q}^{22}-4\,{q}^{20}-9\,{q}^{18}-16\,{q}^{16}-25\,{q}^{
14}-35\,{q}^{12}-47\,{q}^{10}-59\,{q}^{8}-70\,{q}^{6}-79\,{q}^{4}-85\,
{q}^{2}-87-85\,{q}^{-2}-79\,{q}^{-4}-70\,{q}^{-6}-59\,{q}^{-8}-47\,{q}
^{-10}-35\,{q}^{-12}-25\,{q}^{-14}-16\,{q}^{-16}-9\,{q}^{-18}-4\,{q}^{
-20}-{q}^{-22}
$ \end{center} \vspace{-4ex} \begin{center} $
u_{7}^{4}=-{q}^{22}-3\,{q}^{20}-6\,{q}^{18}-9\,{q}^{16}-11\,{q}^{
14}-13\,{q}^{12}-15\,{q}^{10}-17\,{q}^{8}-20\,{q}^{6}-22\,{q}^{4}-24\,
{q}^{2}-25-24\,{q}^{-2}-22\,{q}^{-4}-20\,{q}^{-6}-17\,{q}^{-8}-15\,{q}
^{-10}-13\,{q}^{-12}-11\,{q}^{-14}-9\,{q}^{-16}-6\,{q}^{-18}-3\,{q}^{-
20}-{q}^{-22}
$ \end{center} \vspace{-4ex} \begin{center} $
u_{7}^{7}={q}^{16}+3\,{q}^{14}+5\,{q}^{12}+6\,{q}^{10}+6\,{q}^{8}+
6\,{q}^{6}+8\,{q}^{4}+10\,{q}^{2}+11+10\,{q}^{-2}+8\,{q}^{-4}+6\,{q}^{
-6}+6\,{q}^{-8}+6\,{q}^{-10}+5\,{q}^{-12}+3\,{q}^{-14}+{q}^{-16}
$ \end{center} \vspace{-4ex} \begin{center} $
u_{7}^{8}=-{q}^{14}-{q}^{12}-{q}^{10}-2\,{q}^{6}-2\,{q}^{4}-3\,{q}
^{2}-2-3\,{q}^{-2}-2\,{q}^{-4}-2\,{q}^{-6}-{q}^{-10}-{q}^{-12}-{q}^{-
14}
$ \end{center} \vspace{-4ex} \begin{center} $
u_{7}^{9}={q}^{18}+4\,{q}^{16}+10\,{q}^{14}+18\,{q}^{12}+25\,{q}^{
10}+31\,{q}^{8}+37\,{q}^{6}+44\,{q}^{4}+50\,{q}^{2}+53+50\,{q}^{-2}+44
\,{q}^{-4}+37\,{q}^{-6}+31\,{q}^{-8}+25\,{q}^{-10}+18\,{q}^{-12}+10\,{
q}^{-14}+4\,{q}^{-16}+{q}^{-18}
$ \end{center} \vspace{-4ex} \begin{center} $
u_{7}^{9}={q}^{18}+4\,{q}^{16}+10\,{q}^{14}+18\,{q}^{12}+25\,{q}^{
10}+31\,{q}^{8}+37\,{q}^{6}+44\,{q}^{4}+50\,{q}^{2}+53+50\,{q}^{-2}+44
\,{q}^{-4}+37\,{q}^{-6}+31\,{q}^{-8}+25\,{q}^{-10}+18\,{q}^{-12}+10\,{
q}^{-14}+4\,{q}^{-16}+{q}^{-18}
$ \end{center} \vspace{-4ex} \begin{center} $
u_{8}^{2}=-{q}^{10}-2\,{q}^{8}-3\,{q}^{6}-3\,{q}^{4}-2\,{q}^{2}-1-
2\,{q}^{-2}-3\,{q}^{-4}-3\,{q}^{-6}-2\,{q}^{-8}-{q}^{-10}
$ \end{center} \vspace{-4ex} \begin{center} $
u_{8}^{3}={q}^{8}+{q}^{2}-1+{q}^{-2}+{q}^{-8}
$ \end{center} \vspace{-4ex} \begin{center} $
u_{8}^{4}=-{q}^{22}-3\,{q}^{20}-7\,{q}^{18}-14\,{q}^{16}-24\,{q}^{
14}-36\,{q}^{12}-48\,{q}^{10}-59\,{q}^{8}-70\,{q}^{6}-79\,{q}^{4}-86\,
{q}^{2}-89-86\,{q}^{-2}-79\,{q}^{-4}-70\,{q}^{-6}-59\,{q}^{-8}-48\,{q}
^{-10}-36\,{q}^{-12}-24\,{q}^{-14}-14\,{q}^{-16}-7\,{q}^{-18}-3\,{q}^{
-20}-{q}^{-22}
$ \end{center} \vspace{-4ex} \begin{center} $
u_{8}^{5}=-{q}^{6}-{q}^{4}-1-{q}^{-4}-{q}^{-6}
$ \end{center} \vspace{-4ex} \begin{center} $
u_{8}^{6}={q}^{8}+{q}^{6}+2\,{q}^{4}+3\,{q}^{2}+3+3\,{q}^{-2}+2\,{
q}^{-4}+{q}^{-6}+{q}^{-8}
$ \end{center} \vspace{-4ex} \begin{center} $
u_{8}^{7}=-{q}^{14}-{q}^{12}-{q}^{10}-2\,{q}^{6}-2\,{q}^{4}-3\,{q}
^{2}-2-3\,{q}^{-2}-2\,{q}^{-4}-2\,{q}^{-6}-{q}^{-10}-{q}^{-12}-{q}^{-
14}
$ \end{center} \vspace{-4ex} \begin{center} $
u_{8}^{8}=-53-{q}^{26}-4\,{q}^{22}-6\,{q}^{20}-16\,{q}^{-16}-41\,{
q}^{-8}-6\,{q}^{-20}-{q}^{-26}-2\,{q}^{24}-41\,{q}^{8}-52\,{q}^{-2}-52
\,{q}^{2}-4\,{q}^{-22}-20\,{q}^{14}-26\,{q}^{12}-11\,{q}^{18}-33\,{q}^
{10}-16\,{q}^{16}-2\,{q}^{-24}-11\,{q}^{-18}-20\,{q}^{-14}-26\,{q}^{-
12}-33\,{q}^{-10}-44\,{q}^{6}-44\,{q}^{-6}-47\,{q}^{4}-47\,{q}^{-4}
$ \end{center} \vspace{-4ex} \begin{center} $
u_{8}^{9}=-{q}^{14}-{q}^{12}-2\,{q}^{10}-2\,{q}^{6}-2\,{q}^{4}-2\,
{q}^{2}-2-2\,{q}^{-2}-2\,{q}^{-4}-2\,{q}^{-6}-2\,{q}^{-10}-{q}^{-12}-{
q}^{-14}
$ \end{center} \vspace{-4ex} \begin{center} $
u_{8}^{10}=-{q}^{14}-{q}^{12}-2\,{q}^{10}-2\,{q}^{8}-3\,{q}^{6}-3
\,{q}^{4}-2\,{q}^{2}-3-2\,{q}^{-2}-3\,{q}^{-4}-3\,{q}^{-6}-2\,{q}^{-8}
-2\,{q}^{-10}-{q}^{-12}-{q}^{-14}
$ \end{center} \vspace{-4ex} \begin{center} $
u_{8}^{11}=-{q}^{10}-{q}^{8}-2\,{q}^{6}-2\,{q}^{4}-2\,{q}^{2}-3-2
\,{q}^{-2}-2\,{q}^{-4}-2\,{q}^{-6}-{q}^{-8}-{q}^{-10}
$ \end{center} \vspace{-4ex} \begin{center} $
u_{8}^{12}={q}^{14}+2\,{q}^{12}+3\,{q}^{10}+4\,{q}^{8}+6\,{q}^{6}+
7\,{q}^{4}+7\,{q}^{2}+7+7\,{q}^{-2}+7\,{q}^{-4}+6\,{q}^{-6}+4\,{q}^{-8
}+3\,{q}^{-10}+2\,{q}^{-12}+{q}^{-14}
$ \end{center} \vspace{-4ex} \begin{center} $
u_{9}^{2}=443+{q}^{28}+5\,{q}^{26}+29\,{q}^{22}+51\,{q}^{20}+118\,
{q}^{-16}+325\,{q}^{-8}+51\,{q}^{-20}+5\,{q}^{-26}+{q}^{-28}+14\,{q}^{
24}+325\,{q}^{8}+435\,{q}^{-2}+435\,{q}^{2}+29\,{q}^{-22}+164\,{q}^{14
}+216\,{q}^{12}+80\,{q}^{18}+271\,{q}^{10}+118\,{q}^{16}+14\,{q}^{-24}
+80\,{q}^{-18}+164\,{q}^{-14}+216\,{q}^{-12}+271\,{q}^{-10}+373\,{q}^{
6}+373\,{q}^{-6}+411\,{q}^{4}+411\,{q}^{-4}
$ \end{center} \vspace{-4ex} \begin{center} $
u_{9}^{3}={q}^{18}+{q}^{16}+{q}^{14}-{q}^{10}-2\,{q}^{8}-2\,{q}^{6
}-3\,{q}^{4}-5\,{q}^{2}-5-5\,{q}^{-2}-3\,{q}^{-4}-2\,{q}^{-6}-2\,{q}^{
-8}-{q}^{-10}+{q}^{-14}+{q}^{-16}+{q}^{-18}
$ \end{center} \vspace{-4ex} \begin{center} $
u_{9}^{4}=-5+{q}^{28}+4\,{q}^{26}+15\,{q}^{22}+19\,{q}^{20}+22\,{q
}^{-16}+12\,{q}^{-8}+19\,{q}^{-20}+4\,{q}^{-26}+{q}^{-28}+9\,{q}^{24}+
12\,{q}^{8}-4\,{q}^{-2}-4\,{q}^{2}+15\,{q}^{-22}+21\,{q}^{14}+20\,{q}^
{12}+21\,{q}^{18}+17\,{q}^{10}+22\,{q}^{16}+9\,{q}^{-24}+21\,{q}^{-18}
+21\,{q}^{-14}+20\,{q}^{-12}+17\,{q}^{-10}+7\,{q}^{6}+7\,{q}^{-6}+{q}^
{4}+{q}^{-4}
$ \end{center} \vspace{-4ex} \begin{center} $
u_{9}^{5}=-{q}^{20}-3\,{q}^{18}-5\,{q}^{16}-6\,{q}^{14}-6\,{q}^{12
}-6\,{q}^{10}-7\,{q}^{8}-9\,{q}^{6}-9\,{q}^{4}-8\,{q}^{2}-7-8\,{q}^{-2
}-9\,{q}^{-4}-9\,{q}^{-6}-7\,{q}^{-8}-6\,{q}^{-10}-6\,{q}^{-12}-6\,{q}
^{-14}-5\,{q}^{-16}-3\,{q}^{-18}-{q}^{-20}
$ \end{center} \vspace{-4ex} \begin{center} $
u_{9}^{7}={q}^{18}+4\,{q}^{16}+10\,{q}^{14}+18\,{q}^{12}+25\,{q}^{
10}+31\,{q}^{8}+37\,{q}^{6}+44\,{q}^{4}+50\,{q}^{2}+53+50\,{q}^{-2}+44
\,{q}^{-4}+37\,{q}^{-6}+31\,{q}^{-8}+25\,{q}^{-10}+18\,{q}^{-12}+10\,{
q}^{-14}+4\,{q}^{-16}+{q}^{-18}
$ \end{center} \vspace{-4ex} \begin{center} $
u_{9}^{8}=-{q}^{14}-{q}^{12}-2\,{q}^{10}-2\,{q}^{6}-2\,{q}^{4}-2\,
{q}^{2}-2-2\,{q}^{-2}-2\,{q}^{-4}-2\,{q}^{-6}-2\,{q}^{-10}-{q}^{-12}-{
q}^{-14}
$ \end{center} \vspace{-4ex} \begin{center} $
u_{9}^{9}=845+{q}^{26}+18\,{q}^{22}+47\,{q}^{20}+164\,{q}^{-16}+
577\,{q}^{-8}+47\,{q}^{-20}+{q}^{-26}+5\,{q}^{24}+577\,{q}^{8}+828\,{q
}^{-2}+828\,{q}^{2}+18\,{q}^{-22}+248\,{q}^{14}+346\,{q}^{12}+96\,{q}^
{18}+458\,{q}^{10}+164\,{q}^{16}+5\,{q}^{-24}+96\,{q}^{-18}+248\,{q}^{
-14}+346\,{q}^{-12}+458\,{q}^{-10}+688\,{q}^{6}+688\,{q}^{-6}+775\,{q}
^{4}+775\,{q}^{-4}
$ \end{center} \vspace{-4ex} \begin{center} $
u_{9}^{10}=-{q}^{8}-{q}^{6}-2\,{q}^{4}-{q}^{2}-{q}^{-2}-2\,{q}^{-4
}-{q}^{-6}-{q}^{-8}
$ \end{center} \vspace{-4ex} \begin{center} $
u_{10}^{2}={q}^{16}+2\,{q}^{14}+2\,{q}^{12}+4\,{q}^{10}+6\,{q}^{8}
+7\,{q}^{6}+7\,{q}^{4}+8\,{q}^{2}+8+8\,{q}^{-2}+7\,{q}^{-4}+7\,{q}^{-6
}+6\,{q}^{-8}+4\,{q}^{-10}+2\,{q}^{-12}+2\,{q}^{-14}+{q}^{-16}
$ \end{center} \vspace{-4ex} \begin{center} $
u_{10}^{3}=-{q}^{6}+{q}^{2}-1+{q}^{-2}-{q}^{-6}
$ \end{center} \vspace{-4ex} \begin{center} $
u_{10}^{4}={q}^{22}+{q}^{20}-2\,{q}^{16}-5\,{q}^{14}-9\,{q}^{12}-
15\,{q}^{10}-22\,{q}^{8}-26\,{q}^{6}-30\,{q}^{4}-32\,{q}^{2}-32-32\,{q
}^{-2}-30\,{q}^{-4}-26\,{q}^{-6}-22\,{q}^{-8}-15\,{q}^{-10}-9\,{q}^{-
12}-5\,{q}^{-14}-2\,{q}^{-16}+{q}^{-20}+{q}^{-22}
$ \end{center} \vspace{-4ex} \begin{center} $
u_{10}^{5}={q}^{8}+2\,{q}^{6}+{q}^{4}+{q}^{2}+1+{q}^{-2}+{q}^{-4}+
2\,{q}^{-6}+{q}^{-8}
$ \end{center} \vspace{-4ex} \begin{center} $
u_{10}^{6}=-{q}^{6}+{q}^{2}+1+{q}^{-2}-{q}^{-6}
$ \end{center} \vspace{-4ex} \begin{center} $
u_{10}^{8}=-{q}^{14}-{q}^{12}-2\,{q}^{10}-2\,{q}^{8}-3\,{q}^{6}-3
\,{q}^{4}-2\,{q}^{2}-3-2\,{q}^{-2}-3\,{q}^{-4}-3\,{q}^{-6}-2\,{q}^{-8}
-2\,{q}^{-10}-{q}^{-12}-{q}^{-14}
$ \end{center} \vspace{-4ex} \begin{center} $
u_{10}^{9}=-{q}^{8}-{q}^{6}-2\,{q}^{4}-{q}^{2}-{q}^{-2}-2\,{q}^{-4
}-{q}^{-6}-{q}^{-8}
$ \end{center} \vspace{-4ex} \begin{center} $
u_{10}^{10}=-{q}^{8}-{q}^{4}-{q}^{2}+1-{q}^{-2}-{q}^{-4}-{q}^{-8}
$ \end{center} \vspace{-4ex} \begin{center} $
u_{11}^{2}={q}^{12}+{q}^{6}+{q}^{2}+1+{q}^{-2}+{q}^{-6}+{q}^{-12}
$ \end{center} \vspace{-4ex} \begin{center} $
u_{11}^{4}=-{q}^{16}-2\,{q}^{14}-4\,{q}^{12}-5\,{q}^{10}-6\,{q}^{8
}-8\,{q}^{6}-9\,{q}^{4}-10\,{q}^{2}-11-10\,{q}^{-2}-9\,{q}^{-4}-8\,{q}
^{-6}-6\,{q}^{-8}-5\,{q}^{-10}-4\,{q}^{-12}-2\,{q}^{-14}-{q}^{-16}
$ \end{center} \vspace{-4ex} \begin{center} $
u_{11}^{8}={q}^{10}+{q}^{8}+2\,{q}^{6}+2\,{q}^{4}+2\,{q}^{2}+3+2\,
{q}^{-2}+2\,{q}^{-4}+2\,{q}^{-6}+{q}^{-8}+{q}^{-10}
$ \end{center} \vspace{-4ex} \begin{center} $
u_{12}^{2}=-{q}^{16}-{q}^{14}-2\,{q}^{12}-3\,{q}^{10}-4\,{q}^{8}-5
\,{q}^{6}-5\,{q}^{4}-6\,{q}^{2}-5-6\,{q}^{-2}-5\,{q}^{-4}-5\,{q}^{-6}-
4\,{q}^{-8}-3\,{q}^{-10}-2\,{q}^{-12}-{q}^{-14}-{q}^{-16}
$ \end{center} \vspace{-4ex} \begin{center} $
u_{12}^{4}={q}^{18}+2\,{q}^{16}+4\,{q}^{14}+6\,{q}^{12}+9\,{q}^{10
}+12\,{q}^{8}+14\,{q}^{6}+16\,{q}^{4}+17\,{q}^{2}+17+17\,{q}^{-2}+16\,
{q}^{-4}+14\,{q}^{-6}+12\,{q}^{-8}+9\,{q}^{-10}+6\,{q}^{-12}+4\,{q}^{-
14}+2\,{q}^{-16}+{q}^{-18}
$ \end{center} \vspace{-4ex} \begin{center} $
u_{12}^{8}=-{q}^{14}-2\,{q}^{12}-3\,{q}^{10}-4\,{q}^{8}-6\,{q}^{6}
-7\,{q}^{4}-7\,{q}^{2}-7-7\,{q}^{-2}-7\,{q}^{-4}-6\,{q}^{-6}-4\,{q}^{-
8}-3\,{q}^{-10}-2\,{q}^{-12}-{q}^{-14}
$ \end{center} \vspace{-4ex} \begin{center} $
u_{1}^{1}=u_{1}^{7}=u_1^{10}=u_{1}^{11}=u_{6}^{1}=u_{7}^{1}=u_{10}^{1}=u_{3}^{6}=u_{3}^{8}=u_{3}^{11}=u_{5}^{12}=u_{6}^{7}=u_{6}^{12}=u_{7}^{3}=u_{7}^{10}=u_{7}^{11}=u_{9}^{6}=u_{9}^{11}=u_{10}^{7}=u_{10}^{11}=u_{11}^{3}=u_{11}^{7}=u_{11}^{12}=u_{12}^{3}=u_{12}^{6}=u_{12}^{7}=u_{12}^{9}=u_{12}^{10}=u_{12}^{11}=-1
$
\end{center} \vspace{-4ex} \begin{center} $
u_{7}^{12}=u_{10}^{12}=u_{12}^{12}=u_1^{12}=u_{11}^{11}=u_{6}^{11}=u_{11}^{6}=u_{11}^{7}=u_{11}^{10}=u_{7}^{6}=u_{3}^{10}=u_{7}^{5}=u_{9}^{12}=u_{11}^{5}=u_{11}^{9}=u_{12}^{5}=1, \ \ \ u_{3}^{12}=0
$ \end{center}

\end{document}